\documentclass[12pt,a4paper]{amsart}
\usepackage{amsmath}
\usepackage{paralist}
\usepackage{graphics} 
\usepackage{epsfig} 
\usepackage{graphicx}  
\usepackage{epstopdf}
\usepackage{float}
\usepackage[utf8]{inputenc} 

\usepackage{amssymb}
\usepackage{amsfonts}
\usepackage{amsthm}
\usepackage{ulem}
\usepackage{xcolor}
\usepackage{wrapfig}
\usepackage{subcaption}
\usepackage{tikz}
\usepackage{comment}
\newcommand{\unit}[1]{\mathrm{#1}}
\newcommand{\jmp}[1]{\,[\![#1]\!]}

\newcommand{\ux}{\underline{x}}
\newcommand{\un}{\underline{n}}
\newcommand{\uu}{\underline{u}}
\newcommand{\uf}{\underline{f}}
\newcommand{\uE}{\underline{E}}
\newcommand{\uD}{\underline{D}}
\newcommand{\uF}{\underline{F}}
\newcommand{\uJ}{\underline{J}}

\title[A Theoretical Study of Aqueous Humor Secretion]
{A Theoretical Study of Aqueous Humor Secretion Based on a Continuum Model Coupling Electrochemical and Fluid-Dynamical Transmembrane Mechanisms}

\address{$^{1}$ Universit\'e de Strasbourg, CNRS, IRMA UMR 7501, Strasbourg, France}
\address{$^{2}$ Dipartimento di Matematica, Politecnico di Milano,  
				 Piazza L. da Vinci 32, 20133 Milano, Italy}
\address{$^{3}$ Ospedale Luigi Sacco, via G.~B.~Grassi 74, 20157 Milano, Italy}
\address{$^{4}$ Department of Electrical Engineering and Computer Science, 
College of Engineering, University of Missouri, 201 Naka Hall, Columbia, MO 65211}
\address{$^{5}$ Eugene and Marilyn Glick Eye Institute, 
Indiana University School of Medicine, 1160 W Michigan St, 
Indianapolis, IN 46202, USA}

\author[L.~Sala, A.~G.~Mauri, R.~Sacco, D.~Messenio, G.~Guidoboni, A.~Harris]
{Lorenzo Sala$^{1}$ \and Aurelio Giancarlo Mauri$^{2}$  \and Riccardo Sacco$^{2}$ 
\and Dario Messenio$^{3}$ \and Giovanna Guidoboni$^{4}$ \and Alon Harris$^{5}$}
\email{lorenzo.sala@etu.unistra.fr}
\email{aureliogiancarlo.mauri@polimi.it}
\email{segreteria.oculistica@hsacco.it, dmessenio@virgilio.it}
\email{guidobonig@missouri.edu}
\email{alharris@indiana.edu}
\email{riccardo.sacco@polimi.it}

\begin{document}

\date{\today}

\begin{abstract}
Intraocular pressure, resulting from the balance of aqueous humor (AH) production and drainage, is the only approved treatable risk factor in glaucoma. AH production is determined by the concurrent function of ionic pumps and aquaporins in the ciliary processes but their individual contribution is difficult to characterize experimentally. In this work, we propose a novel unified modeling and computational framework for the finite element simulation of the role of
the main ionic pumps involved in AH secretion, namely, the sodium-potassium pump, the calcium-sodium pump, the anion channel and the hydrogenate-sodium pump. The theoretical model 
is developed at the cellular scale and is based on the coupling between electrochemical 
and fluid-dynamical transmembrane mechanisms characterized by 
a novel description of the electric pressure exerted by the ions on the intrachannel fluid that 
includes electrochemical and osmotic corrections. Considering a realistic
geometry of the ionic pumps, the proposed model is demonstrated to correctly predict 
their functionality as a function of (1) the permanent electric charge density over the channel 
pump surface; (2) the osmotic gradient coefficient; (3) the stoichiometric ratio between the ionic pump currents enforced at the inlet and outlet sections of the channel. In particular, theoretical predictions of the 
transepithelial membrane potential for each simulated pump/channel allow us to perform a first 
significant model comparison with experimental data for monkeys. This is a significant step 
for future multidisciplinary studies on the action of molecules on AH production.

\end{abstract}

\maketitle


\vspace{5mm}
\noindent

\section{Introduction}\label{sec:introduction}
The flow of aqueous humor (AH) and its regulation play an important role in ocular physiology by
contributing 
to control the level of intraocular pressure (IOP)~\cite{Moses1987,Kiel1998}. 
Elevated IOP is the only approved treatable risk factor in glaucoma,
an optic neuropathy characterized by a multifactorial aetiology with a progressive degeneration of retinal ganglion cells that ultimately leads to irreversible vision loss~\cite{glaucoma_state_of_the_art_GG,glaucoma_state_of_the_art}).
Currently, glaucoma affects more than 60 million people worldwide  and is estimated to reach almost 80 millions by 2020~\cite{Quigley2006}.
IOP can be lowered via hypotonizing eye drops and/or surgical treatment,and it can be 
shown that reducing IOP by $1 \; \unit{mmHg}$ has the effect of reducing
the risk of glaucoma progression and subsequent vision loss by 10\%~\cite{heijl2002reduction}.  

Several classes of IOP-lowering medications are available for use in patients with glaucoma, including prostaglandin analogues, beta-blockers, carbonic anhydrase inhibitors, alpha-2-adrenergic agonists,  in fixed and variable associations, while newer classes are still in clinical trials, such as the rho kinase inhibitors \cite{goldhagen2012elevated,honjo2001effects,rao2007rho}. 
All currently available IOP-lowering agents function by altering AH production or drainage.  
However, differences in drug efficacy have been observed among patients that cannot be completely explained without a clear understanding of the mechanisms regulating AH flow. Motivated by this need, in this work we focus on AH production and we propose a mathematical approach to model and simulate the contribution of ionic pumps and channels to determine AH flow.

\begin{figure}[h!]
\centering
\includegraphics[width=0.8\textwidth]{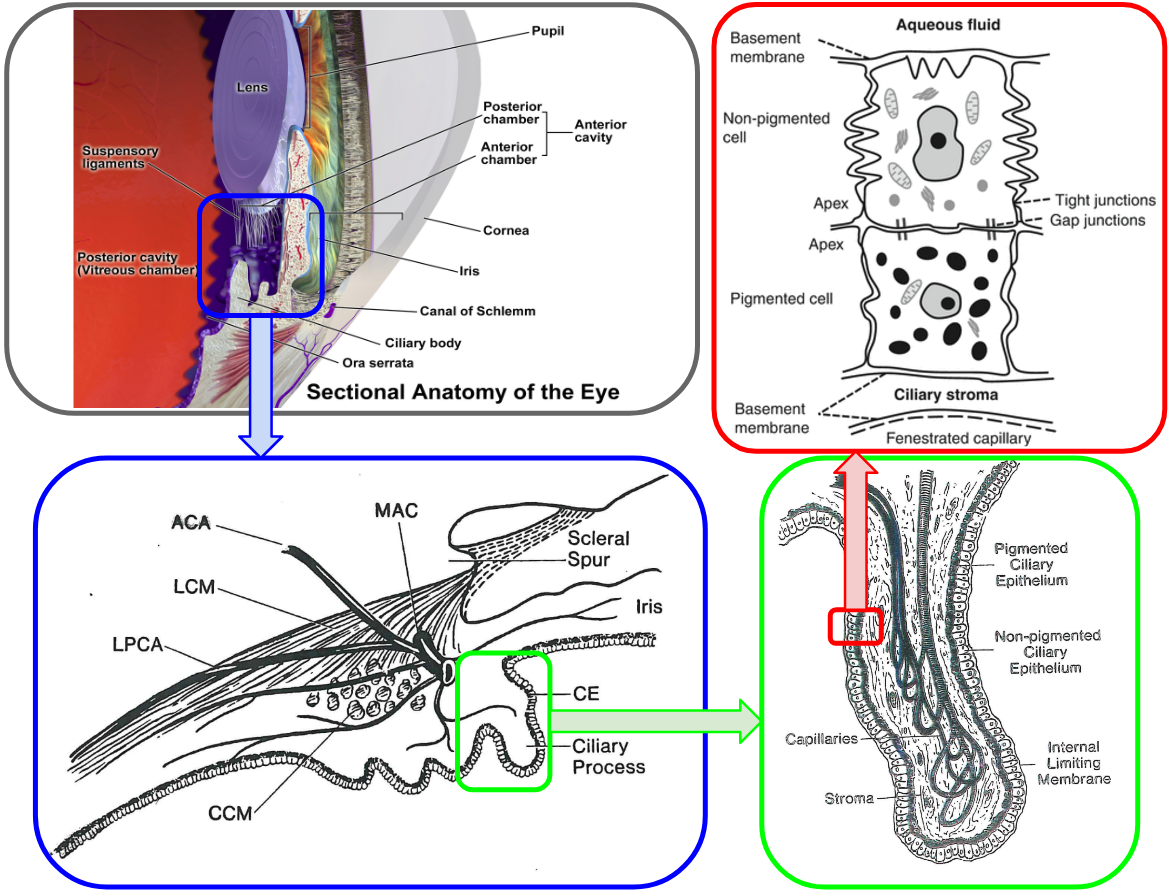}
\caption{ \newline \underline{Left top panel:} anatomy of the eye and of the structures involved in aqueous humor production and regulation. (Blausen.com staff(2014). "Medical gallery of Blausen Medical 2014". WikiJournal of Medicine 1(2). DOI:10.15347/wjm/2014.010. ISSN 2002-4436) \\
	\underline{Left bottom panel:} MAC: major arterial circle; ACA: anterior ciliary arteries; LPCA: long posterior ciliary artery; LCM: longitudinal (fibers) ciliary muscle; CCM: circular (fibers) ciliary muscle; CE: ciliary 
	epithelium (figure reproduced from \cite{shields1982study}) \\
	\underline{Right bottom panel:} A ciliary process is composed by capillaries, stroma and two layer of epithelium (inner, pigmented and outer, nonpigmented) (figure reproduced from \cite{shields1982study}) \\
	\underline{Right top panel:} the two-layer structure of the ciliary epithelium (figure reproduced from {\tt https://entokey.com/wp-content/uploads/2016/}).} 
\label{fig:AH}
\end{figure}

The production of AH takes places in the ciliary processes within the ciliary body, where clear liquid
 flows across the ciliary epithelium, a two-layered structure composed by an inner nonpigmented layer, representing the continuation of retinal pigmented epithelium, and an external nonpigmented layer, representing the continuation of the retina~\cite{Goel2010}, as illustrated in Figure~\ref{fig:AH}.

Three main mechanisms are involved in the production of AH: 
\textit{(i)} convective delivery of fluid and metabolic components via the ciliary circulation;
\textit{(ii)} ultrafiltration and diffusion of fluid and metabolic components  across the epithelial cells driven by gradients in hydrostatic pressure, oncotic pressure and metabolite concentrations;
\textit{(iii)} active secretion into the posterior chamber driven by increased ionic concentrations within
the basolateral space between nonpigmented epithelial cells. 

In this work we focus on the third mechanism, henceforth referred to as \textit{AH secretion}, which is 
responsible for approximately $80-90\%$ of the whole AH production process~\cite{Gabelt2003,Mark2009}. 
More precisely, we aim at modeling
 the selective movement of anions and cations across the membrane of the nonpigmented epithelial cells,
the resulting gradient of ion and solute concentrations across the membrane, and 
 the induced fluid egression into the posterior chamber.
 
The proposed simulation of AH secretion presents many challenges from the modeling viewpoint. 
Existing references concerning AH secretion are
primarily based on lumped parameter models~that provide a systemic view of AH flow \cite{Brubaker1989,Kiel2011,Szopos2016}, but
do not reproduce the detailed phenomena occurring at the level of single ionic pumps. Detailed models based on the Velocity-Extended Poisson-Nerst-Planck (VE-PNP) system have been utilized to simulate electro-kinetic 
flows~\cite{Aluru2005,Bazant2008,Bazant2015}, but different models for the volumetric force coupling electrochemical and fluid-dynamical mechanisms have been proposed~\cite{Stratton,Rosenfeld2,Roth2,EisenbergJChemP2010}, 
thereby raising the question of which one, if any, is the most appropriate for the application at hand.  In addition, some of the most important parameters in the VE-PNP model, such as the 
concentration of ions within the ionic channel, the fixed charge on the channel lateral surface, 
the osmotic diffusive parameter, cannot be easily accessed experimentally and so 
are not readily available in the literature.  
In the pilot investigation conducted in~\cite{Mauri2016}, we explored
the feasibility of utilizing a VE-PNP model to simulate the sodium-potassium pump (Na$^+$-K$^+$) 
within the nonpigmented epithelial cells of the eye. 
However, several other ionic pumps and channels are involved in AH secretion
~\cite{Kiel1998}, and have not yet been modeled in the context of AH flow.

The present work aims at extending the modeling and simulation treatment of~\cite{Mauri2016} 
through the development of
a unified framework capable of simulating AH secretion by including
the four main ionic pumps and channels involved in the process, namely, the 
calcium-sodium pump (Ca$^{++}$-Na$^+$), the carbonic anhydrase-activated  anion channel (Cl$^-$ - HCO$_3^-$) and sodium-hydrogen exchanger (Na$^+$-H$^+$). 
The computational structure used in the numerical
simulations is based on the adoption of 
(1) a temporal semi-discretization with the Backward Euler method; (2) a Picard iteration to successively 
solve the equation system at each discrete time level; and (3) a spatial discretization of each differential 
subproblem obtained from system decoupling using the Galerkin Finite Element Method.
Numerical simulations are utilized to: 
\textit{(i)} compare how and to what extent different modeling choices for the volumetric coupling force 
affect the resulting transmembrane potential, stoichiometric ratio and intrachannel fluid velocity; and 
\textit{(ii)} characterize the correct boundary conditions and the value of the permanent 
electric charge density on the channel lateral surface that allow us to predict a biophysically
reasonable behavior of ionic pumps and channels in realistic geometries. 
Overall, this work provides the first systematic investigation of VE-PNP models in the context of AH secretion 
and paves the way to future studies on biochemical, pharmacological and
therapeutical aspects of AH flow regulation.

The paper is organized as follows. 
Section~\ref{sec:AH} provides a brief functional description of the main features pertaining
the Na$^+$-K$^+$, Ca$^{++}$-Na$^+$ and Cl$^-$ - HCO$_3^-$ pumps and the Na$^+$-H$^+$ channel. 
The VE-PNP system is described in Section~\ref{sec:model}
and the mathematical model for the volume force density 
in the right-hand side of the linear momentum balance equation for the 
intrachannel fluid is described in Section~\ref{sec:forzanti}.
The numerical discretization of the VE-PNP model equations is discussed 
in Section~\ref{sec:discretization}, whereas simulation results of
the effect of volumetric forces 
and permanent electric charge density are presented in Sections~\ref{sec:simulations_1} 
and~\ref{sec:simulations_2}, respectively. Model limitations,
conclusions and future perspectives are outlined in Section~\ref{sec:final_conclusions}.

\section{Ionic pumps and channels in AH secretion}
\label{sec:AH}
In this section we provide a short description of the main ionic pumps and channels that are involved in the process of AH secretion.
We refer to~\cite{Hille2001} for an overview of ionic channels in cellular biology and to~\cite{KeenerSneyd1998} 
for a mathematical analysis of ionic pumps in cellular biology.

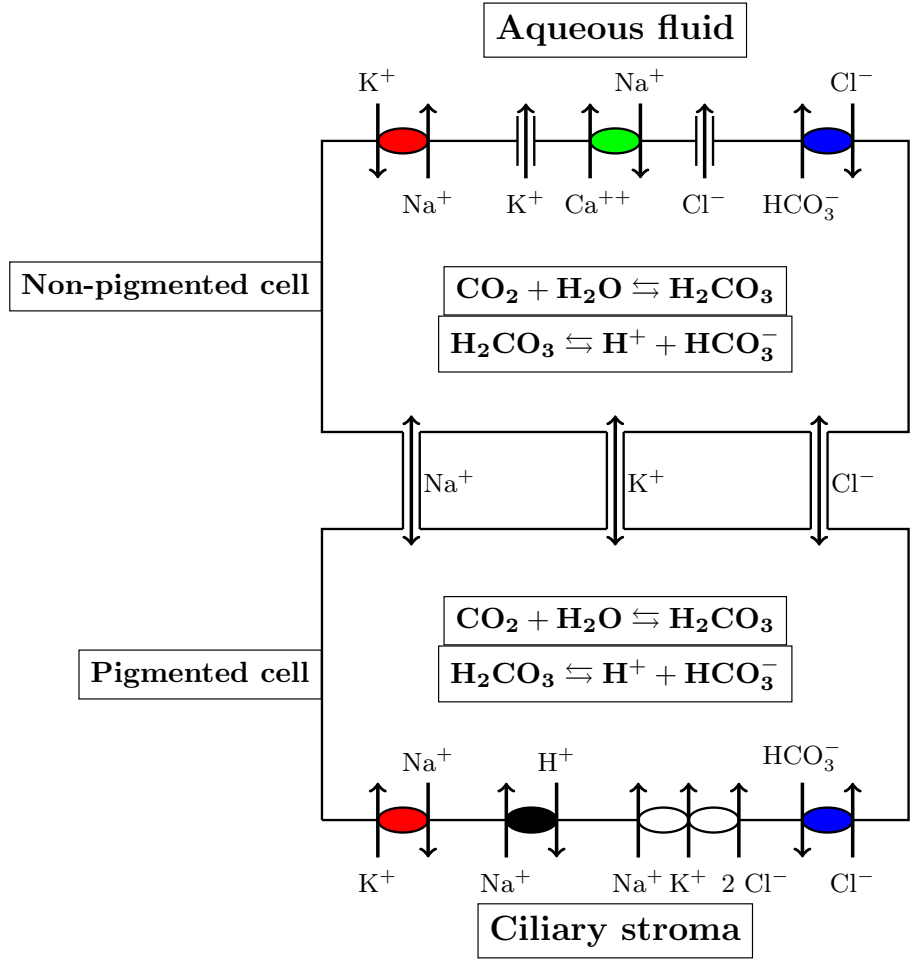
\begin{figure}[h!]
		\centering
		\resizebox{0.8\textwidth}{!}{\begin{tikzpicture}
\def\xl{7}
\def\H{\xl+\xl/6}
\def\h{\xl/2}

\draw[thick] (0-\xl/2,0) -- (0+\xl/2,0)
	-- (\xl/2,\h) -- (\xl/3+0.2,\h);
\draw[thick] (0+\xl/3,\h) -- (0.1,\h);
\draw[thick] (0-\xl/3,\h) -- (-0.1,\h);
\draw[thick] (0-\xl/3-0.2,\h) -- (0-\xl/2,\h)
	-- (0-\xl/2,0);
\draw[thick] (-\xl/3-0.2,\h+\xl/6) -- (-\xl/2,\h+\xl/6) 
	-- (-\xl/2,\H) -- (-\xl/6,\H);
\draw[thick] (-\xl/6+0.2,\H) -- (\xl/6-0.2,\H);
\draw[thick] (\xl/6,\H) -- (\xl/2,\H)
	-- (0+\xl/2,\h+\xl/6) -- (\xl/3+0.2,\h+\xl/6);
\draw[thick] (-\xl/3,\h+\xl/6) -- (-0.1,\h+\xl/6);
\draw[thick] (\xl/3,\h+\xl/6) -- (0.1,\h+\xl/6);
\draw[thick] (0-\xl/3,\h) -- (0-\xl/3,\h+\xl/6);
\draw[thick] (0-\xl/3-0.2,\h) -- (0-\xl/3-0.2,\h+\xl/6);
\draw[thick] (0+\xl/3,\h) -- (0+\xl/3,\h+\xl/6);
\draw[thick] (0+\xl/3+0.2,\h) -- (0+\xl/3+0.2,\h+\xl/6);
\draw[thick] (0.1,\h) -- (0.1,\h+\xl/6);
\draw[thick] (-0.1,\h) -- (-0.1,\h+\xl/6);

\node[below,draw] at (0,-1) {\textbf{Ciliary stroma}};
\node[above,draw] at (0,\H+1) {\textbf{Aqueous fluid}};
\node[left,draw] at (-\xl/2,\H-\h/2) {\footnotesize \textbf{Non-pigmented cell}};
\node[left,draw] at (-\xl/2,\h/2) {\footnotesize \textbf{Pigmented cell}};

\draw[thick,fill=red] (-\xl/3-0.2,\H) ellipse (0.3 and 0.15);
\draw[thick,fill=blue] (\xl/3+0.2,\H) ellipse (0.3 and 0.15);
\draw[thick,fill=green] (0,\H) ellipse (0.3 and 0.15);
\draw[thick] (\xl/6,\H-0.3) -- (\xl/6,\H+0.3);
\draw[thick] (\xl/6-0.2,\H-0.3) -- (\xl/6-0.2,\H+0.3);
\draw[thick] (-\xl/6,\H-0.3) -- (-\xl/6,\H+0.3);
\draw[thick] (-\xl/6+0.2,\H-0.3) -- (-\xl/6+0.2,\H+0.3);
\draw[thick,fill=red] (-\xl/3-0.2,0) ellipse (0.3 and 0.15);
\draw[thick,fill=blue] (\xl/3+0.2,0) ellipse (0.3 and 0.15);
\draw[thick,fill=black] (-\xl/7,0) ellipse (0.3 and 0.15);
\draw[thick,fill=white] (\xl/8+0.3,0) ellipse (0.3 and 0.15);
\draw[thick,fill=white] (\xl/8-0.3,0) ellipse (0.3 and 0.15);

\draw[very thick,->] (\xl/6-0.1,\H-0.45) -- (\xl/6-0.1,\H+0.45);
	\node[below] at (\xl/6-0.1,\H-0.45) {\scriptsize Cl$^-$};
\draw[very thick,->] (-\xl/6+0.1,\H-0.45) -- (-\xl/6+0.1,\H+0.45);
	\node[below] at (-\xl/6+0.1,\H-0.45) {\scriptsize K$^+$};
\draw[very thick,<-] (-\xl/3-0.5,\H-0.45) -- (-\xl/3-0.5,\H+0.45);
	\node[above] at (-\xl/3-0.5,\H+0.45) {\scriptsize K$^+$};
\draw[very thick,->] (-\xl/3+0.1,\H-0.45) -- (-\xl/3+0.1,\H+0.45);
	\node[below] at (-\xl/3+0.1,\H-0.45) {\scriptsize Na$^+$};
\draw[very thick,->] (\xl/3-0.1,\H-0.45) -- (\xl/3-0.1,\H+0.45);
	\node[below] at (\xl/3-0.1,\H-0.45) {\scriptsize HCO$_3^-$};
\draw[very thick,<-] (\xl/3+0.5,\H-0.45) -- (\xl/3+0.5,\H+0.45);
	\node[above] at (\xl/3+0.5,\H+0.45) {\scriptsize Cl$^-$};
\draw[very thick,<-] (0.3,\H-0.45) -- (0.3,\H+0.45);
	\node[above] at (0.3,\H+0.45) {\scriptsize Na$^+$};
\draw[very thick,->] (-0.3,\H-0.45) -- (-0.3,\H+0.45);
	\node[below] at (-0.2,\H-0.45) {\scriptsize Ca$^{++}$};

\draw[very thick,<->] (-\xl/3-0.1,\h-0.2) -- (-\xl/3-0.1,\h+\xl/6+0.2);
	\node[right] at (-\xl/3-0.1,\h+\xl/12) {\scriptsize Na$^+$};
\draw[very thick,<->] (\xl/3+0.1,\h-0.2) -- (\xl/3+0.1,\h+\xl/6+0.2);
	\node[right] at (\xl/3+0.1,\h+\xl/12) {\scriptsize Cl$^-$};
\draw[very thick,<->] (0,\h-0.2) -- (0,\h+\xl/6+0.2);
	\node[right] at (0,\h+\xl/12) {\scriptsize K$^+$};
\draw[very thick,->] (-\xl/3-0.5,-0.45) -- (-\xl/3-0.5,0.45);
	\node[below] at (-\xl/3-0.5,-0.45) {\scriptsize K$^+$};
\draw[very thick,<-] (-\xl/3+0.1,-0.45) -- (-\xl/3+0.1,0.45);
	\node[above] at (-\xl/3+0.1,0.45) {\scriptsize Na$^+$};
\draw[very thick,->] (-\xl/7-0.3,-0.45) -- (-\xl/7-0.3,0.45);
	\node[below] at (-\xl/7-0.3,-0.45) {\scriptsize Na$^+$};
\draw[very thick,<-] (-\xl/7+0.3,-0.45) -- (-\xl/7+0.3,0.45);
	\node[above] at (-\xl/7+0.3,0.45) {\scriptsize H$^+$};
\draw[very thick,->] (\xl/8-0.6,-0.45) -- (\xl/8-0.6,0.45);
	\node[below] at (\xl/8-0.6,-0.45) {\scriptsize Na$^+$};
\draw[very thick,->] (\xl/8,-0.45) -- (\xl/8,0.45);
	\node[below] at (\xl/8,-0.45) {\scriptsize K$^+$};
\draw[very thick,->] (\xl/8+0.6,-0.45) -- (\xl/8+0.6,0.45);
	\node[below] at (\xl/8+0.8,-0.45) {\scriptsize 2 Cl$^-$};
\draw[very thick,<-] (\xl/3-0.1,-0.45) -- (\xl/3-0.1,0.45);
	\node[above] at (\xl/3-0.1,0.45) {\scriptsize HCO$_3^-$};
\draw[very thick,->] (\xl/3+0.5,-0.45) -- (\xl/3+0.5,0.45);
	\node[below] at (\xl/3+0.5,-0.45) {\scriptsize Cl$^-$};

\node[draw,above] at (0,3*\h/5) {\footnotesize $\mathbf{CO_2 + H_2O \leftrightarrows H_2CO_3}$};
\node[draw,below] at (0,3*\h/5) {\footnotesize $\mathbf{H_2CO_3 \leftrightarrows H^+ + HCO_3^-}$};
\node[draw,above] at (0,\H-3*\h/5) {\footnotesize $\mathbf{CO_2 + H_2O \leftrightarrows H_2CO_3}$};
\node[draw,below] at (0,\H-3*\h/5) {\footnotesize $\mathbf{H_2CO_3 \leftrightarrows H^+ + HCO_3^-}$};
\end{tikzpicture}}
		\caption{Schematic diagram of ionic pumps and ionic channels located on the lipid membrane in 
		the nonpigmented epithelial cells of the ciliary body of the eye. Aqueous humor is produced by
		the active secretion of fluid through the ionic channels during their activity.
		This figure is inspired by Figure~9 of~\cite{Shahidullah2011}. \\
		}\label{fig:ionic_channel}
\end{figure}

\subsection{The sodium-potassion ATPasi pump.}
\begin{figure}[ht!]
	\centering
	\begin{tikzpicture}
		\draw[thick] (-1,1.75) arc (-180:-90:1 and 0.75);
		\draw[thick] (0,1)	-- (3,1);
		\draw[thick] (4,1.75) arc (0:-90:1 and 0.75);
		\draw[thick] (-1,-0.75) arc (180:90:1 and 0.75);
		\draw[thick] (0,0)	-- (3,0);
		\draw[thick] (4,-0.75) arc (0:90:1 and 0.75);
		\node[] at (-4,0.5) {intracellular}; 
		\node[] at (-4,0.1) {space}; 
		\node[] at (6.5,0.5) {extracellular};
		\node[] at (6.5,0.1) {space}; 
		\node[] at (1.5,1.5) {membrane};		
		\draw[fill] (2.8,0.7) -- (3.3,0.7) -- (3.3,0.6) -- (2.8,0.6)
		-- (2.8,0.55) -- (2.7,0.65) -- (2.8,0.75) -- (2.8,0.7) -- cycle;
		\draw[fill] (2.8,0.4) -- (3.3,0.4) -- (3.3,0.3) -- (2.8,0.3)
		-- (2.8,0.25) -- (2.7,0.35) -- (2.8,0.45) -- (2.8,0.4) -- cycle;		
		\node[] at (3.7,0.55) {$\mathbf{K}^+$};
		\draw [fill](0.2,0.75) -- (-0.3,0.75) -- (-0.3,0.85) -- (0.2,0.85)
		-- (0.2,0.9) -- (0.3,0.8) -- (0.2,0.7) -- (0.2,0.75) -- cycle;
		\draw[fill] (0.2,0.45) -- (-0.3,0.45) -- (-0.3,0.55) -- (0.2,0.55)		
		-- (0.2,0.6) -- (0.3,0.5) -- (0.2,0.4) -- (0.2,0.45) -- cycle;
		\draw[fill] (0.2,0.15) -- (-0.3,0.15) -- (-0.3,0.25) -- (0.2,0.25)
		-- (0.2,0.3) -- (0.3,0.2) -- (0.2,0.1) -- (0.2,0.15) -- cycle;
		\node[] at (-0.9,0.55) {$\mathbf{Na}^+$};
\end{tikzpicture}
	\caption{Schematic representation of the Na$^+$-K$^+$ pump. The stoichiometric ratio is $3:2$, 
	since there is an outflux of three Na$^+$ ions and an influx of two K$^+$ ions.}
	\label{KNapump}			
\end{figure}
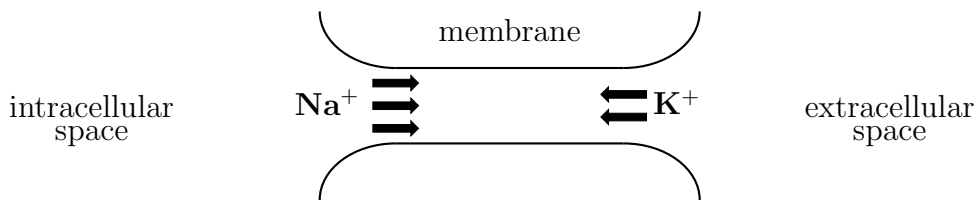

This pump plays a fundamental role in cellular biology as it is present in the membrane of every cell in the human body.
The enzyme ATPasi, located either in pigmented or nonpigmented ciliary epithelium, causes the hydrolysis of one molecule of ATP 
and produces the necessary energy to expel three Na$^+$ ions, while allowing two K$^+$ ions to enter. 
This process is not electrically neutral as it entails an outflux of three positive charged particles of sodium
and an influx of only two positive charged particles of potassium. 
The ion outflux and influx are schematically represented in Figure~\ref{KNapump}.

\subsection{The calcium-sodium ATPasi pump.}
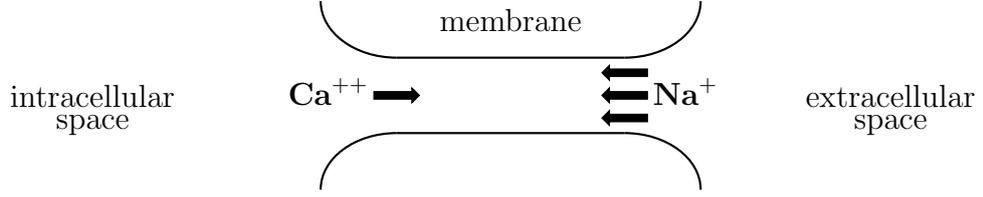
\begin{figure}[ht!]
	\centering
	\begin{tikzpicture}
		\draw[thick] (-1,1.75) arc (-180:-90:1 and 0.75);
		\draw[thick] (0,1)	-- (3,1);
		\draw[thick] (4,1.75) arc (0:-90:1 and 0.75);
		\draw[thick] (-1,-0.75) arc (180:90:1 and 0.75);
		\draw[thick] (0,0)	-- (3,0);
		\draw[thick] (4,-0.75) arc (0:90:1 and 0.75);
		\node[] at (-4,0.5) {intracellular}; 
		\node[] at (-4,0.1) {space}; 
		\node[] at (6.5,0.5) {extracellular};
		\node[] at (6.5,0.1) {space}; 
		\node[] at (1.5,1.5) {membrane};		
		\draw[fill] (2.8,0.85) -- (3.3,0.85) -- (3.3,0.75) -- (2.8,0.75)
		-- (2.8,0.7) -- (2.7,0.8) -- (2.8,0.9) -- (2.8,0.85) -- cycle;
		\draw[fill] (2.8,0.55) -- (3.3,0.55) -- (3.3,0.45) -- (2.8,0.45)
		-- (2.8,0.4) -- (2.7,0.5) -- (2.8,0.6) -- (2.8,0.55) -- cycle;
		\draw[fill] (2.8,0.25) -- (3.3,0.25) -- (3.3,0.15) -- (2.8,0.15)
		-- (2.8,0.1) -- (2.7,0.2) -- (2.8,0.3) -- (2.8,0.25) -- cycle;		
		\node[] at (3.8,0.55) {$\mathbf{Na}^+$};
		\draw[fill] (0.2,0.45) -- (-0.3,0.45) -- (-0.3,0.55) -- (0.2,0.55)		
		-- (0.2,0.6) -- (0.3,0.5) -- (0.2,0.4) -- (0.2,0.45) -- cycle;
		\node[] at (-0.9,0.55) {$\mathbf{Ca}^{++}$};
\end{tikzpicture}		
	\caption{Schematic representation of the Ca$^{++}$-Na$^+$ pump.  The stoichiometric ratio is $3:1$, since there is an influx of three Na$^+$ ions and an outflux of one Ca$^{++}$ ion.}
	\label{CaNapump}
\end{figure}

This pump is activated when calcium accumulates inside the cell above a certain threshold, that is usually around $1\; \unit{mM}$. 
This pump entails the influx of three Na$^+$ ions and an outflux of one Ca$^{++}$ ion. 
As in the previous case, this process is not electrically neutral as three positive sodium ions enter the cell whereas
only one positive calcium ion exits the cell. 
The ion outflux and influx are schematically represented in Figure~\ref{CaNapump}.

\subsection{The carbonic anhydrase-activated anionic pump.} 
\begin{figure}[ht!]
	\centering
				\centering
		\begin{tikzpicture}
		\node[] at (-4,-0.8) {intracellular}; 
		\node[] at (-4,-1.2) {space}; 
		\node[] at (6.5,-0.8) {extracellular};
		\node[] at (6.5,-1.2) {space}; 
		\node[] at (1.5,1.5) {membrane};
		\node[] at (1.5,-1.5) {membrane};
		\draw[thick] (-1,1.75) arc (-180:-90:1 and 0.75);
		\draw[thick] (0,1)	-- (3,1);
		\draw[thick] (4,1.75) arc (0:-90:1 and 0.75);
		\draw[thick] (-1,-0.75) arc (180:90:1 and 0.75);
		\draw[thick] (0,0)	-- (3,0);
		\draw[thick] (4,-0.75) arc (0:90:1 and 0.75);		
		\draw[fill] (2.8,0.55) -- (3.3,0.55) -- (3.3,0.45) -- (2.8,0.45)
		-- (2.8,0.4) -- (2.7,0.5) -- (2.8,0.6) -- (2.8,0.55) -- cycle;
		\node[] at (3.75,0.55) {$\mathbf{Cl}^-$};
		\draw[fill] (0.2,0.45) -- (-0.3,0.45) -- (-0.3,0.55) -- (0.2,0.55)		
		-- (0.2,0.6) -- (0.3,0.5) -- (0.2,0.4) -- (0.2,0.45) -- cycle;
		\node[] at (-0.9,0.55) {$\mathbf{HCO}_3^-$};
		\draw[thick] (-1,-1.25) arc (-180:-90:1 and 0.75);
		\draw[thick] (0,-2)	-- (3,-2);
		\draw[thick] (4,-1.25) arc (0:-90:1 and 0.75);
		\draw[thick] (-1,-3.75) arc (180:90:1 and 0.75);
		\draw[thick] (0,-3)	-- (3,-3);
		\draw[thick] (4,-3.75) arc (0:90:1 and 0.75);		
		\draw[fill] (2.8,-2.45) -- (3.3,-2.45) -- (3.3,-2.55) -- (2.8,-2.55)
		-- (2.8,-2.6) -- (2.7,-2.5) -- (2.8,-2.4) -- (2.8,-2.45) -- cycle;
		\node[] at (3.8,-2.45) {$\mathbf{Na}^+$};
		\draw[fill] (0.2,-2.55) -- (-0.3,-2.55) -- (-0.3,-2.45) -- (0.2,-2.45)		
		-- (0.2,-2.4) -- (0.3,-2.5) -- (0.2,-2.6) -- (0.2,-2.55) -- cycle;
		\node[] at (-0.9,-2.45) {$\mathbf{H}^+$};
		\end{tikzpicture}
	\caption{Schematic representation of the carbonic anhydrase-activated anionic pump (top panel) and 
	of the Na$^+$ and H$^+$ exchanger (bottom panel).
	The stoichiometric ratio of the anionic pump  is $1:1$, since there is an influx of one Cl$^-$ ion 
	and an outflux of one HCO$_3^{-}$ ion. The stoichiometric ratio of the  Na$^+$ and H$^+$ exchanger 
	is also 1:1, since there is an influx of one Na$^+$ ion and an outflux of one H$^+$ ion.}
	\label{carbonanydrasepump}
\end{figure}
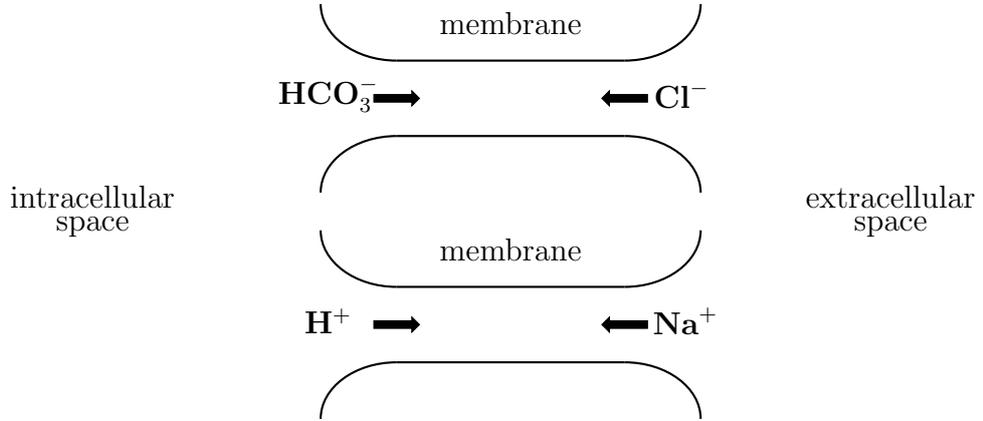

This pump involves the movement of negative ions. 
Carbonic anhydrase is an enzyme that mediates the transport of bicarbonate across the ciliary epithelium to maintain 
the homeostatic balance of carbonate across the cell membrane. 
More precisely, carbonic anhydrase favors the splitting of one molecule of H$_2$CO$_3$ into a positive H$^+$ ion and a 
negative HCO$_3^-$ ion. Then, the HCO$_3^-$ ion exits the cell through the anion channel with an exchange of a chlorine ion Cl$^-$ 
that enters the cell. 
The balance of this pump is electrically neutral because for every negative charged HCO$_3^-$ leaving the cell there is a negative 
charged Cl$^-$ ion entering the cell~\cite{Wistrand1951}.
The ion outflux and influx are schematically represented in the top panel of Figure~\ref{carbonanydrasepump}.

\subsection{{\rm Na}$^+$ and {\rm H}$^+$ exchanger.}
This pump is strictly correlated with the activity of the carbonic anhydrase-activated anionic pump previously described. 
A positive H$^+$ ion resulting from the splitting reaction of one molecule of H$_2$CO$_3$ exits the cell with an exchange 
of one Na$^+$ ion entering the cell. Thus, this pump is electrically neutral. 
The ion outflux and influx are schematically represented in the bottom panel of Figure~\ref{carbonanydrasepump}.

\section{The mathematical model}\label{sec:model}
In this section we illustrate the system of partial differential equations (PDEs) constituting the mathematical model at the cellular 
scale level of ionic channel and pumps that activate the AH secretion.
We refer to~\cite{Art_equations} for a detailed discussion of the analytical properties of the model equations 
and of the numerical methods used for their discretization, and to~\cite{Mauri2016} for preliminary results on 
the adoption of the model in the study of the role of bicarbonate ion to correctly determine the electrostatic potential 
drop across the cellular membrane at the level of eye transepitelium.

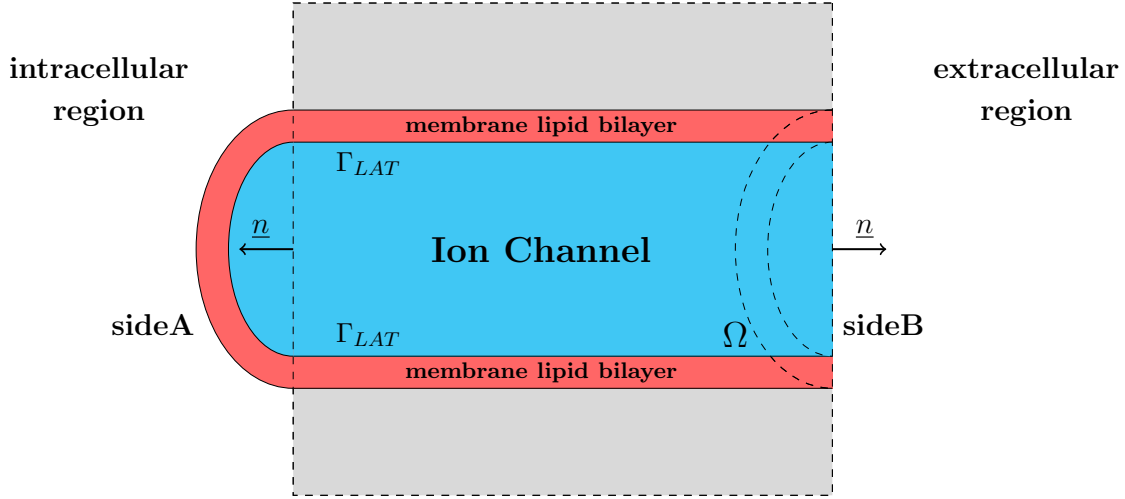
\begin{figure}[ht!]
\centering
\resizebox{\textwidth}{!}{\begin{tikzpicture}[scale=1.5]
\def\l{1.3}
\draw[dashed, fill=white!85!black] (3,\l+1) -- (3,-\l-1)
		-- (-2,-\l-1) -- (-2,\l+1) -- cycle;
\fill[white!40!red] (-2,\l) -- (3,\l) -- (3,1) -- (-2,1)-- cycle;
\fill[white!40!red] (-2,-\l) -- (3,-\l) -- (3,-1) -- (-2,-1)-- cycle;
\draw[dashed,fill=white!40!red] (3,\l) arc (-270:-90:0.9 and \l); 
\fill[white!40!cyan] (-2,1) -- (3,1) -- (3,-1) -- (-2,-1) -- cycle;
\draw (-2,1) -- (3,1);
\draw (-2,-1) -- (3,-1);
\draw[fill=white!40!red] (-2,\l) arc (-270:-90:0.9 and \l); 
\draw (-2,\l) -- (3,\l);
\draw (-2,-\l) -- (3,-\l);
\draw[fill=white!40!cyan] (-2,1) arc (-270:-90:0.6 and 1); 
\draw[dashed,fill=white!40!cyan] (3,1) arc (-270:-90:0.6 and 1);
\draw[dashed] (3,\l) arc (-270:-90:0.9 and \l);
\draw[thick,->] (3,0) -- (3.5,0);
\draw[thick,->] (-2,0) -- (-2.5,0);
\draw[dashed] (3,\l+1) -- (3,-\l-1)
		-- (-2,-\l-1) -- (-2,\l+1) -- cycle;
\node[] at (2.1,-0.8) {\large $\Omega$};
\node[] at (0.3,0) {\large \textbf{Ion Channel}};
\node[] at (0.3,1.15) {\scriptsize \textbf{membrane lipid bilayer}};
\node[] at (0.3,-1.15) {\scriptsize \textbf{membrane lipid bilayer}};
\node[above] at (-3.8,1.5) {\textbf{intracellular}};
\node[below] at (-3.8,1.5) {\textbf{region}};
\node[above] at (4.8,1.5) {\textbf{extracellular}};
\node[below] at (4.8,1.5) {\textbf{region}};
\node[above] at (3.3,0) {$\un$};
\node[above] at (-2.3,0) {$\un$};
\node[] at (-3.3,-0.7) {\textbf{sideA}};
\node[right] at (3,-0.7) {\textbf{sideB}};
\node[below] at (-1.3,1) {\small $\Gamma_{LAT}$};
\node[above] at (-1.3,-1) {\small $\Gamma_{LAT}$};
\end{tikzpicture}}
\caption{Cross-section of a simplified ionic channel geometry. 
The lipid membrane bilayer is represented in brown color. 
The ion channel is represented in cyan color. 
The portions of the boundary $\partial \Omega$ are labeled as $sideA$ (intracellular side), $sideB$ (extracellular side) 
and $\Gamma_{LAT}$ (lateral surface). }
\label{fig:Omega_2D}
\end{figure}

The geometrical setting that we consider henceforth is the computational domain $\Omega$ illustrated in Figure~\ref{fig:Omega_2D} 
representing a cross-section of a simplified ion channel geometry. 
Such representation includes any of the ionic pumps or ionic channels described in Section~\ref{sec:AH}, 
which are located on the lipid bilayer constituting the membrane of the nonpigmented epithelial cells of the ciliary body 
of the eye schematically depicted in Figure~\ref{fig:AH} (right top panel).
We indicate by $\partial \Omega$ the boundary of $\Omega$, by $\un$ the outward unit normal vector on $\partial \Omega$ and by 
$sideA$, $sideB$ and $\Gamma_{LAT}$ the intracellular surface, the extracellular surface and the lateral boundary, respectively, 
in such a way that $\partial \Omega = sideA \cup sideB \cup \Gamma_{LAT}$. 
For a given starting time $t_0$ and a given observational time window $T_{obs}$, we set $I_T:= (t_0, t_0+T_{obs})$ and 
we denote by $Q_T:= \Omega \times I_T$ the space-time cylinder in which we study the spatial and temporal evolution of the process of
AH secretion at the cellular scale level.
To clarify the physical foundation of the cellular scale model object of the present article, 
we assume that the following {strongly coupled mechanisms} concur to determine AH secretion:
\begin{itemize}
	\item \emph{electric field formation}: this mechanism is determined by the mutual interaction among ions 
	in the intrachannel {fluid} and their interaction with the permanent 
	{electric charge density} distributed on the surface of the 
	channel structure. Mathematically, the mechanism is described by the Poisson equation, 
	supplied by appropriate boundary conditions at the inlet and outlet sections of the channel and on its external surface;
	\item \emph{ion motion}: this mechanism is determined by the superposition of a diffusion process driven by ion concentration 
	gradients along the channel and of a drift process driven by the force exerted by the electric field on each ion charged 
	particle.
	Mathematically, the mechanism is described by the Nernst-Planck equations, supplied by appropriate initial conditions 
	inside the channel and by appropriate boundary conditions at the inlet and outlet sections of the channel and on its 
	external surface;
	\item \emph{fluid motion}: this mechanism is determined by the volume force density that is exerted by the charged 
	ion particles because of their motion inside the channel.
	Mathematically, the mechanism is described by the time-dependent Stokes equations, supplied by appropriate initial 
	conditions inside the channel and by appropriate boundary conditions at the inlet and outlet sections of the channel
	and on its external surface.
\end{itemize} 

\noindent
We refer to the resulting mathematical model as \textit{Velocity-Extended Poisson-Nernst-Planck} system (VE-PNP)~\cite{rubinstein,Jerome2002,Schmuck,Sacco2009}. 
The VE-PNP system can be derived by the application of the following physical laws~\cite{Art_equations}:
\textit{(i)}  mass balance for each of the $M$ chemical species included in the system \eqref{eq:PNP};
\textit{(ii)}  linear momentum balance for each chemical species \eqref{eq:PNP_flux};
\textit{(iii)}  electrical charge conservation \eqref{eq:PNP_Poisson};
\textit{(iv)}  mass balance {for intrachannel fluid} \eqref{eq:stokes_incomp};
\textit{(v)}  linear momentum balance {for intrachannel fluid} \eqref{eq:stokes_flux}.
Ultimately, the system consists of the following set of PDEs to be solved in $Q_T$:

\begin{subequations}\label{eq:VE_PNP_system}
\begin{flalign}
&\frac{\partial n_i}{\partial t} +  \text{div}\,\uf_i = 0 & \forall i= 1,\dots,M & \label{eq:PNP}\\
&\uf_i = \frac{z_i}{|z_i|} \mu_i n_i \uE - D_i \nabla n_i + \boxed{n_i \,\uu} & \forall i= 1,\dots,M & \label{eq:PNP_flux}\\
&\text{div} \uD = q \sum_{i=1}^{M} z_i n_i + q \rho_{fixed} 
& \label{eq:PNP_Poisson}\\
& \uD = \epsilon_f \uE & \label{eq:displacement} \\
& \uE = -\nabla \varphi & \label{eq:PNP_potential} \\[2mm]
& \text{div}\,\uu = 0 & \label{eq:stokes_incomp} \\
&\rho_f \frac{\partial \uu}{\partial t} = 
\text{div }\underline{\underline{\sigma}}(\uu,p) + 
\boxed{\uF_{ion}} & \label{eq:stokes_flux}  \\
&\underline{\underline{\sigma}} (\uu,p) = 
2\mu_f \underline{\underline{\epsilon}}(\uu) - p\,\underline{\underline{\delta}} 
& \label{eq:stress_tensor} \\
& \underline{\underline{\epsilon}}(\uu) =
\underline{\underline{\nabla}}_s \uu = 
\dfrac{1}{2} (\nabla \uu + (\nabla \uu)^T). & \label{eq:strain_rate}
\end{flalign}
\end{subequations}

The equation set~\eqref{eq:VE_PNP_system} comprises two main blocks. 
Equations~\eqref{eq:PNP}-~\eqref{eq:PNP_potential} constitute the Poisson-Nernst-Planck (PNP) block whereas equations~\eqref{eq:stokes_incomp}-~\eqref{eq:strain_rate} constitute the Stokes block.
As far as the PNP block is concerned, $M$ is the number of ionic species, $\uf_i$ denotes the ion particle flux $[\unit{cm^{-2} \,s^{-1}}]$, $n_i$ is the ion concentration $[\unit{cm^{-3}}]$, $\mu_i$ and $D_i$ are the ion electrical mobility 
$[\unit{cm^2\,V\,s^{-1}}]$ and diffusivity $[\unit{cm^2\,s^{-1}}]$ and $T$ is the absolute temperature $[\unit{K}]$. 
The mobility $\mu_i$ is proportional to the diffusivity $D_i$ through the Einstein relation
\begin{align}
& D_i = \dfrac{K_B T}{q |z_i|} \mu_i \qquad i=1, \ldots, M, & \label{eq:einstein}
\end{align}
where $q$ is the electron charge $[\unit{C}]$ and $K_B$ is the Boltzmann constant $[\unit{cm^2 g s^{-2} K^{-1}}]$.
In Eq. \eqref{eq:PNP_Poisson} and Eq. \eqref{eq:PNP_potential}, $\uD$ is the electric displacement $[\unit{C \, cm^{-2}}]$, $\uE$ is the electric field $[\unit{V\, cm^{-1}}]$, $\varphi$ is the electric potential $[\unit{V}]$ and $\epsilon_f$ 
is the electrolyte fluid dielectric permittivity $[\unit{F\,cm^{-1}}]$. 
The quantity $z_i$ is the valence of the $i$-th ion ($z_i>0$ for cations, $z_i<0$ for anions) whereas $\rho_{fixed}$ is the permanent {electric} 
charge density $[\unit{C \, cm^{-3}}]$. 
We define the ionic current density $\uJ_i$ $[\unit{A\,cm^{-2}}]$ of each ion species as 
\begin{align}
& \uJ_i = q z_i \, \uf_i \qquad i=1, \ldots, M. & \label{eq:current_density}
\end{align}
As far as the Stokes block is concerned, $\uu$ is the fluid velocity $[\unit{cm\,s^{-1}}]$ and fluid incompressibility is expressed 
by Eq. (\ref{eq:stokes_incomp}).
Eq. \eqref{eq:stress_tensor} and Eq. \eqref{eq:strain_rate} are the constitutive laws for the stress tensor $[\unit{dyne \; cm^{-2}}]$ 
and the strain rate $[\unit{s^{-1}}]$ respectively, where $p$ denotes the fluid pressure $[\unit{dyne \; cm^{-2}}]$, $\mu_f$ 
is the fluid viscosity $[\unit{g \, cm^{-1} \, s^{-1}}]$, $\rho_f$ is the fluid mass density $[\unit{g \, cm^{-3}}]$ 
and $\underline{\underline{\delta}}$ the second-order identity tensor of dimension 3. 
Since the focus of the article is the investigation of the role of electrochemical forces on the secretion of AH across the 
nonpigmented epithelial cells in the ciliary body of the eye, we assume that the temperature $T$
of the intrachannel fluid is constant and equal to the value $T_0 = 293.75 \unit{K}$.
The boxed terms in Eq.~\eqref{eq:PNP_flux} and Eq.~\eqref{eq:stokes_flux} are the contributions that introduce the coupling between 
the PNP block of system~\eqref{eq:VE_PNP_system} and the Stokes block of system~\eqref{eq:VE_PNP_system}.
In particular, 
\begin{align}
& \uF_{ion} = \sum_{i=1}^{M}\uF_i & \label{eq:volume_force_density}
\end{align}
expresses the volume force density $[\unit{dyne \; cm^{-3}}]$ exerted by the ionic charges on the 
intrachannel fluid, the quantities $\uF_i$ representing the contribution to the volume force density
given by each ion species, $i=1, \ldots, M$.

The equation system~\eqref{eq:VE_PNP_system} is equipped with the following initial conditions:

\begin{subequations}\label{eq:ics_bcs}
\begin{align}
& n_i(\ux,0) = n_i^0(\ux) \qquad i=1, \ldots, M, \qquad \ux \in 
\Omega, & \label{eq:ics_NP} \\
& \uu(\ux,0) = \uu^0(\ux) \qquad \ux \in 
\Omega, & \label{eq:ic_stokes} 
\end{align}
where $n_i^0$ are given positive functions and $\uu^0$ is a given function.
The initial condition $\varphi^0=\varphi^0(\ux)$, $\ux \in \Omega$, is the solution of the equation 
set~\eqref{eq:PNP_Poisson}-~\eqref{eq:PNP_potential}, under appropriate boundary conditions on $\partial \Omega$, 
having set $n_i=n_i^0$, $i=1, \ldots, M$.
\end{subequations}

The boundary conditions associated with system~\eqref{eq:VE_PNP_system} that are considered in the present article are 
of mixed Dirichlet-Neumann type. Their characterization for each equation in the system 
is specified in each simulation illustrated in Sections~\ref{sec:simulations_1} and~\ref{sec:simulations_2}.

\section{Model for the volume force density}\label{sec:forzanti}
In this section we discuss a general approach to the modeling of the volume force density $\uF_i$ on
the right-hand side in the linear momentum balance equation~\eqref{eq:stokes_flux}. 
$\uF_i$ expresses the contribution from the $i$-th ionic species, $i=1, \ldots, M$, 
to the total volume force density exerted by the ion charged particles on the intrachannel fluid
and is assumed henceforth to be characterized by the following relation
\begin{equation}
\uF_i = q z_i n_i \; \uE_i^{echs} - k_{osm} \, \nabla n_i \qquad i=1, \ldots, M. \label{eq:echsk_force}
\end{equation}
The first term at the right-hand side of~\eqref{eq:echsk_force} 
represents the volume force density due to a generalized electrochemical field $\uE_i^{echs}$ 
whereas the second term represents the volume force density due to 
an osmotic concentration gradient according {to} 
the parameter $k_{osm}$ $[\unit{N \; m}]$ 
(see~\cite{JAEGER1999}). The generalized electrochemical field is {the result of} 
the superposed effect
of passive drift due to the electric field (e), the diffusion mechanism associated with a 
chemical concentration gradient (c) and the particle size exclusion effect associated with 
the hard sphere (hs) theory (see~\cite{HS_1,HS_2}). 
Relation~\eqref{eq:echsk_force} is referred to henceforth as electrochemical model including 
osmotic and size exclusion mechanisms (echsk).

Assuming that $\uE_i^{echs}$ is a gradient field, we have:
\begin{subequations}
\begin{align}
& \uE_i^{echs} = -\nabla \varphi_i^{echs} \qquad i=1, \ldots, M, & \label{eq:gen_electrochemical_field} \\
& \varphi^{echs}_i = \varphi^{ec}_i + \mu_i^{ex} \qquad i=1, \ldots, M, & \label{eq:pot_echs}
\end{align}
where $\varphi^{ec}_i$ is the generalized electrochemical potential of the $i$-th species
\begin{align}
\varphi_i^{ec} = \varphi + \dfrac{V_{th}}{z_i} \ln \left( \dfrac{n_i}{n_{ref}} \right) \qquad i=1, \ldots, M & \label{eq:electrochemical_potential}
\end{align}
and $\mu^{ex}_i$ is the exclusion effect potential of the $i$-th species (see~\cite{breitkopf2016springer})
\begin{align}
& \mu_i^{ex} = - V_{th} \Big[ \ln(1-\frac{4\pi}{3} \sum_{k=1}^M n_k R_k^3) & \nonumber \\
& + 4\pi \frac{R_i (\sum_{k=1}^M n_k R_k^2) + R_i^2 (\sum_{k=1}^M n_k R_k) + 
	\dfrac{1}{3} R_i^3 (\sum_{k=1}^M n_k)}{1-\frac{4\pi}{3} \sum_{k=1}^M n_k R_k^3} & \nonumber \\
& + \dfrac{16 \pi^2}{3} \frac{R_i^3 (\sum_{k=1}^M n_k R_k) 
	(\sum_{k=1}^M n_k R_k^2) + \dfrac{3}{2} R_i^2 \left(\sum_{k=1}^M n_k R_k^2\right)^2}
{\left(1-\frac{4\pi}{3} \sum_{k=1}^M n_k R_k^3\right)^2}   & \nonumber \\
& + \dfrac{64 \pi^3}{9} \frac{R_i^3 
	\left(\sum_{k=1}^M n_k R_k^2\right)^3}{\left(1-\frac{4\pi}{3} 
	\sum_{k=1}^M n_k R_k^3\right)^3} \Big] \qquad i=1, \ldots, M, & \label{eq:hs_muEX}
\end{align}
$n_{ref}$ and $R_i$ being a positive constant $[\unit{cm^{-3}}]$ representing the reference concentration 
in the ionic solution and the ionic radius of the $i$-th ion species, respectively.
\end{subequations}
Relation~\eqref{eq:echsk_force} is indeed a general view of the volume force density from which
{simpler} expressions of $\uF_i$ can be derived. These models 
constitute a hierarchy characterized by an increasing number of approximations and a consequent decreasing
level of physical complexity.

\begin{description}
\item[echs] \textit{Electrochemical model including hard sphere theory.}
This model is derived from~\eqref{eq:echsk_force} by neglecting the contribution of
the osmotic gradient. This includes: the Coulomb electrical force associated with a charge density, 
the chemical gradient and size exclusion mechanisms. 
It is mathematically expressed by the following relation
\begin{equation}
\uF_i = q z_i n_i \; \uE_i^{echs}  \qquad i=1, \ldots, M. \label{eq:echs_force}
\end{equation}

\item[ec] \textit{Electrochemical model.}\label{ch:ec_force}
This is derived from~\eqref{eq:echs_force} neglecting the size exclusion phenomena.
This includes the Coulomb electrical force associated with a charge density 
and the chemical gradient mechanism. 
It is mathematically expressed by the following relations:
\begin{align}
& \uF_i = q z_i n_i \; \uE_i^{ec} \qquad i=1, \ldots, M. & \label{eq:ec_force} \\
& \uE_i^{ec} = -\nabla \varphi_i^{ec} \qquad i=1, \ldots, M. & \label{eq:ec_field}
\end{align}

\item[Stratton] \textit{Stratton model.}\label{ch:std_force}
This model is derived from~\eqref{eq:ec_force} by neglecting the contribution
induced by the chemical gradient. This was originally proposed in~\cite{Stratton} 
and it is widely adopted in the modeling description of electrokinetic phenomena (see~\cite{Aluru2005}). 
It is mathematically expressed by the following relation
\begin{align}
& \uF_i = q z_i n_i \; \uE \qquad i=1, \ldots, M. & \label{eq:standard_force}	
\end{align}

\item[eck]
\textit{Electrochemical model including osmotic force.}\label{ch:eck}
This model is derived from~\eqref{eq:echsk_force} by neglecting the contribution of
the size exclusion phenomena. This model includes the Coulomb electrical 
force associated with a charge density, the chemical and osmotic gradient mechanisms
It is mathematically expressed by the following relation
\begin{equation}
\uF_i = q z_i n_i \; \uE_i^{ec} - k_{osm} \, \nabla n_i \qquad i=1, \ldots, M. \label{eq:eck_force}
\end{equation}

\end{description}

The effect on intrachannel fluid modtion induced by the above force models is analyzed, 
{in the context of the Na$^+$-K$^{-}$ pump}, in Section~\ref{sec:simulations_1} where 
the {predicted} electrolyte fluid velocity is compared with the 
volumetric {force density} $\uF_i$ exerted by the ions on it.

\section{Time advancing, functional iteration and numerical discretization}\label{sec:discretization}
In this section we provide a short description of the algorithm that is used to numerically solve system~\eqref{eq:VE_PNP_system}. 
We refer to~\cite{Art_equations} for more details on the stability and convergence analysis of the adopted methods as well as 
their implementation in the general-purpose C++ modular numerical code MP-FEMOS (Multi-Physics Finite Element Modeling Oriented 
Simulator) that has been developed by some of the authors~\cite{VEPNP,JMI_Reviewed,femos}. 

The VE-PNP model constitutes a nonlinearly coupled system of PDEs of incomplete parabolic type because of the fact that at 
each time level the electrostatic potential $\varphi$ and the electric field $\uE$ must be updated as a function of the ion 
concentrations and of the fixed permanent charge by solving the elliptic Poisson equation~\eqref{eq:PNP_Poisson}. 
In turn, the electric field $\uE$ contributes to determine ion motion through the Nernst-Planck relation~\eqref{eq:PNP_flux} 
and fluid motion through the relation~\eqref{eq:echsk_force} for the volume force density.

To disentangle the various coupling levels that are present in the VE-PNP system, we proceed as follows:

\begin{enumerate}
\item we perform a temporal semi-discretization of the problem by resorting to the Backward Euler (BE) method;
\item we introduce a Picard iteration to successively solve the equation system at each discrete time level;
\item we perform a spatial discretization of each differential subproblem obtained from system 
decoupling using the Galerkin Finite Element Method (GFEM).
\end{enumerate}

The use of the BE method has the twofold advantage that (a) it is unconditionally absolute stable and (b) it is monotone. 
Property (a) allows to take relatively large time steps thus reducing the computational effort to reach steady-state conditions. 
Property (b), combined with an analogous one for the spatial discretization scheme of the Nernst-Planck equations,
ensures that the computed ion concentrations are positive for all discrete time levels.

The use of a Picard iteration has the twofold advantage that (c) it is a decoupled algorithm and (d) a maximum principle is 
satisfied by the solutions of two of the boundary value problems (BVP) obtained from decoupling. 
Property (c) amounts to transforming the nonlinearly coupled system~\eqref{eq:VE_PNP_system} into the successive 
solution of three sets of linear BVPs of reduced size. 
Property (d) implies the existence of an invariant region for the electric potential depending only on the 
boundary data and on the fixed permanent charge and the positivity of the computed ion concentrations.

The use of the GFEM has the twofold advantage that (e) it can easily handle complex geometries and (f) provides an accurate 
and stable numerical approximation of the solution of each BVP obtained from system decoupling. 
Property (e) is implemented through the partition of the domain of biophysical interest into the union of tetrahedral 
elements of variable size. 
Property (f) is implemented through the use of piecewise linear finite elements for the Poisson equation, piecewise 
linear finite elements with exponential fitting stabilization along of mesh edges for the Nernst-Planck equations, and of 
the inf-sup stable Taylor-Hood finite element pair for the Stokes equations. 

\begin{figure}[ht!]
\centering
\includegraphics[width=0.3\textwidth,height=0.6\textwidth]{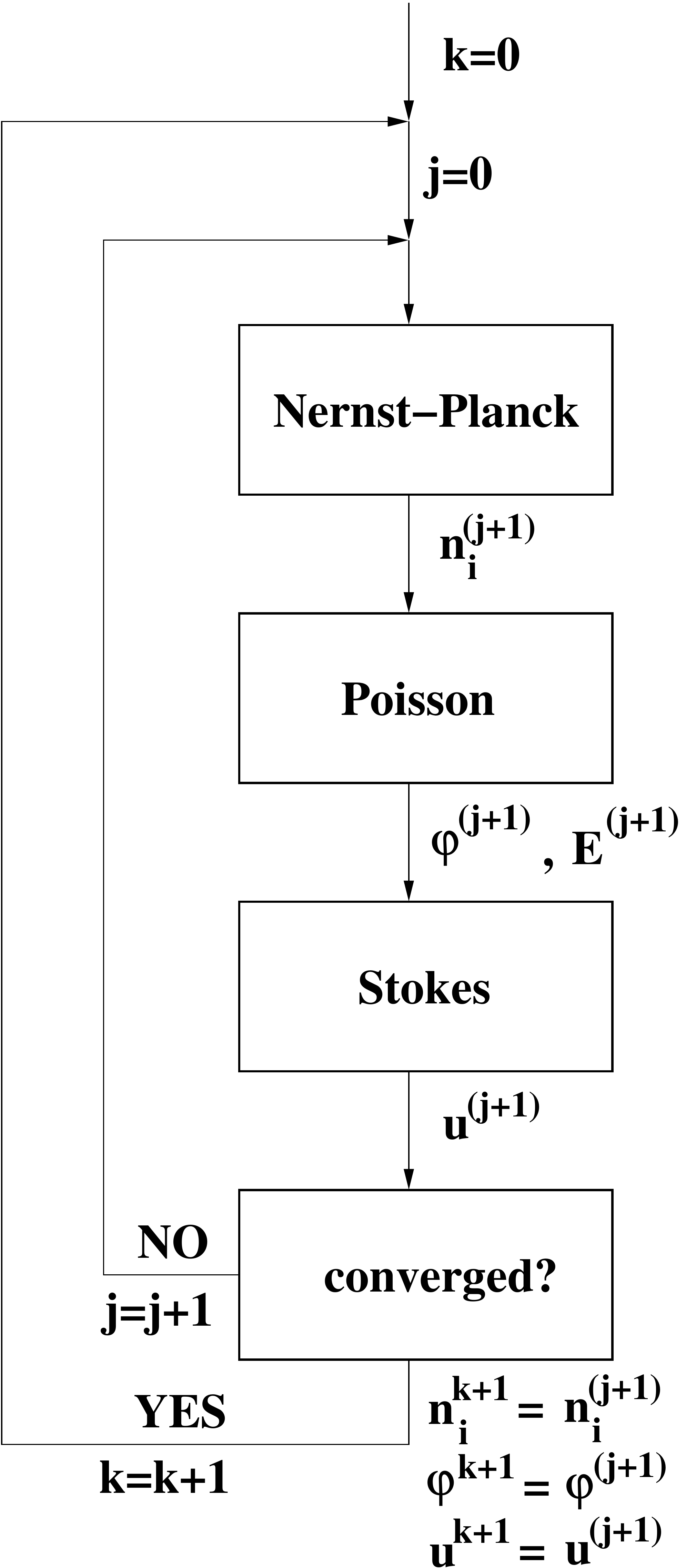}
\caption{Flow-chart of the computational algorithm to solve the VE-PNP system. 
	The nonnegative integer $k$ is the temporal discretization counter. 
	The nonnegative integer $j$ is the Picard iteration counter.}
\label{fig:gummel_algorithm}
\end{figure}

Figure~\ref{fig:gummel_algorithm} illustrates the flow-chart of the temporal semi-discretization for time advancing 
with the BE method and the Picard iteration used to successively solve the three linear equation subsystems.

\section{Comparison of volumetric force models on an 
idealized sodium-potassium pump}\label{sec:simulations_1}
\begin{figure}[h!]
\centering
\includegraphics[width=0.65\textwidth,height=0.25\textwidth]{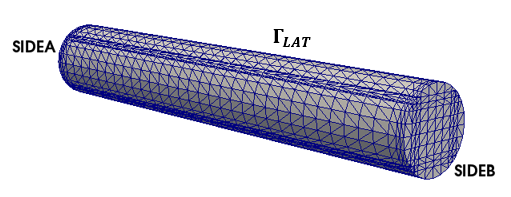}
\caption{Computational domain for the simulation of the Na$^+$-K$^+$ pump.
	The finite element triangulation consists of 37075 tetrahedra.}
\label{fig:cilinder_r04FITTA}
\end{figure}

The aim of this section is to compare the different descriptions 
introduced in Section~\ref{sec:forzanti} of the volume force 
density $\uF_{ion}$ exerted on the intrachannel fluid in the linear momentum balance equation~\eqref{eq:stokes_flux}.
In particular we investigate the biophysical reliability of the various models to describe the functionality of
an {idealized version of the} 
Na$^+$-K$^+$ pump illustrated in Section~\ref{sec:AH} in which
the {simultaneous} presence of Na$^+$, K$^+$, Cl$^-$ and HCO$_3^-$ 
{ionic species is} considered. The analysis criteria are based on the comparison of simulation results with ({\it i}) 
the electrostatic potential drop measured across the transepithelial membrane, ({\it ii}) the theoretical 
stoichiometric ratio $3:2$ and ({\it iii}) the direction of the AH flow from the cell 
into the basolateral space. The ideality of the Na$^+$-K$^+$ pump is represented by the 
geometry {adopted for numerical simulation, consisting of the cylinder 
with axial length $L_{ch} = 5 \unit{nm}$ and radius $R_{ch} = 0.4 \unit{nm}$
shown in Figure~\ref{fig:cilinder_r04FITTA} together with its partition into 37075 tetrahedral finite elements.}
We point out that {the above geometrical setting, despite being a simplified 
approximation of the real structure,} has been {successfully employed 
for biological investigations in~\cite{femos},~\cite{Art_equations} and~\cite{Mauri2016}.}

\subsection{Boundary and initial conditions}
Because of the {complexity of the boundary conditions (BCs) and initial conditions (ICs) 
involved in the simulation of the pump, it is useful to accurately describe them for each 
equation in~\eqref{eq:VE_PNP_system}}.

\subsubsection{Poisson equation}
For all $t \in I_T$, the BCs for the Poisson equation~\eqref{eq:PNP_Poisson}-\eqref{eq:PNP_potential} are:
\begin{subequations}\label{eq:BC_s_Poisson_1}
\begin{align}
& \varphi = 0                     & \qquad \textrm{on } sideA & \label{eq:pot_dirichlet} \\
& \uD \cdot \un = 0 & \qquad \textrm{on } sideB \cup & \label{eq:ef_neumann} \Gamma_{LAT}.
\end{align}
\end{subequations}
{Condition~\eqref{eq:pot_dirichlet} has the scope of introducing a reference value} 
for the calculation of the electrostatic {potential} drop across the cellular membrane.
{Condition~\eqref{eq:ef_neumann} expresses}
the biological fact that the aferomentioned potential drop is caused solely by the ion charge
distribution within the channel because no external bias is applied.

\subsubsection{Nernst-Planck equations}
{The Nernst-Planck equation system allows us to determine the spatial concentration of the 
various ion species inside the channel. The connection between the channel region and the
intra/extracellular sides is made possible by enforcing nonhomogeneous Neumann boundary conditions
that preserve the correct input/output biophysical pump functionality.}
{The boundary and initial conditions for each simulated ion species read 
as follows:}
\begin{description}
\item[K$^+$] 
\begin{subequations}\label{eq:BC_s_pnp_1}
\begin{align}
& \uf_{\textrm{K}^+} \cdot \un = g_{\textrm{K}^+} & \qquad \textrm{on } sideA & \\
& n_{\textrm{K}^+} = \textrm{K}^+_{out} & \qquad \textrm{on } sideB & \\
& \uf_{\textrm{K}^+} \cdot \un = 0 & \qquad \textrm{on } \Gamma_{LAT} &\\
& n_{\textrm{K}^+}(\ux,0) = \textrm{K}^+_0  & \qquad \forall \ux \in \Omega.&
\end{align}

\item[Na$^+$]
\begin{align}
& n_{\textrm{Na}^+} = \textrm{Na}^+_{in} & \qquad \textrm{on } sideA & \\
& \uf_{\textrm{Na}^+} \cdot \un = g_{\textrm{Na}^+} & \qquad \textrm{on } sideB & \\
& \uf_{\textrm{Na}^+} \cdot \un = 0 & \qquad \textrm{on } \Gamma_{LAT}&\\
& n_{\textrm{Na}^+}(\ux,0) = \textrm{Na}^+_0  & \qquad \forall \ux \in \Omega.&
\end{align}

\item[Cl$^-$]
\begin{align}
& n_{\textrm{Cl}^-} = \textrm{Cl}^-_{in} & \qquad \textrm{on } sideA & \\
& n_{\textrm{Cl}^-} = \textrm{Cl}^-_{out} & \qquad \textrm{on } sideB & \\
& \uf_{\textrm{Cl}^-} \cdot \un = 0 & \qquad \textrm{on } \Gamma_{LAT}&\\
& n_{\textrm{Cl}^-}(\ux,0) = \textrm{Cl}^-_0  & \qquad \forall \ux \in \Omega.&
\end{align}

\item[HCO$_3^-$]
\begin{align}
& n_{\textrm{HCO}^-_{3}} = \textrm{HCO}^-_{3,in} & \qquad \textrm{on } sideA & \\
& n_{\textrm{HCO}^-_{3}} = \textrm{HCO}^-_{3,out} & \qquad \textrm{on } sideB & \\
& \uf_{\textrm{HCO}^-_{3}}\cdot \un = 0 & \qquad \textrm{on } \Gamma_{LAT}&\\
& n_{\textrm{HCO}^-_{3}}(\ux,0) = \textrm{HCO}^-_{3,0}  & \qquad \forall \ux \in \Omega.&
\end{align}
\end{subequations}

\end{description}

\noindent

The values of the boundary data for the cations are specified in Table~\ref{tab:Ion+} whereas the values 
of the boundary data for the anions are specified in Table~\ref{tab:Ion-}.
{The boundary values for the ions agree with
the experimental value of the ionic concentrations measured 
in the intracellular and extracellular sides of the {non-pigmented epithelial cells}.} 

\subsubsection{Stokes system}
The calculation of fluid velocity is made possible by {solving} the Stokes system~\eqref{eq:stokes_incomp}-~\eqref{eq:stokes_flux}.
To prescribe a correct biophysical condition of the intrachannel fluid 
{we need to mathematically express that (1) the fluid is adherent to the channel wall; 
(2) no external pressure drop is applied across the channel; (3) 
the fluid is at the rest when the pump is not active. To this purpose, 
the appropriate BCs and ICs read:}
\begin{subequations}\label{eq:BC_s_Stokes_1}
\begin{align}
& \uu = \underline{0}                     & \qquad \textrm{on } \Gamma_{LAT} & \\
& \underline{\underline{{\sigma}}} \:\un = \underline{0} & \qquad \textrm{on } 
sideA \cup sideB &\\
& \uu(\ux,0) = \underline{0}                     & \qquad \forall \ux \in \Omega.&
\end{align}
\end{subequations}

\subsection{Simulation results}
To ease the interpretation of the reported results, we point out that the $Z$ axis coincides with the axial 
direction of the channel and it is positively oriented towards the extracellular region.
Reported data for the vector-valued variables (such as electric field, current densities and velocity) are the $Z$ 
component of the vectors because the other two computed components were comparably negligible. 
We set $\rho_{fixed} = 0 \; [\unit{C \, m^{-3}}]$, $t_0=0 \; [\unit{s}]$ and $T_{obs} = 50 \; [\unit{ns}]$, 
a sufficiently large value to ensure that the simulated system has reached steady-state conditions at $t=T_{obs}$. 
The values of the dielectric permittivity of the intrachannel water fluid $\epsilon_r^{f}$, of the fluid shear viscosity 
$\mu_f$, of the fluid mass density $\rho_f$ and of the diffusion coefficients $D_i$ of each $i$-th ionic species 
involved in the computational tests are reported in Table~\ref{tab:parameters_model_KNa_pump}.
All the figures in the remainder of the section illustrate computed results at $t= T_{obs}$.

\begin{table}[h!]
	\centering
		\begin{tabular}{|l|l|l|}
    \hline
		Model parameter & value & units \\ \hline
		$\epsilon_f$    &  $708.32 \cdot 10^{-10}$     & $[\unit{F cm^{-1}}]$      \\ \hline
		$\mu_f$    &  $10^{-2}$     &   $[\unit{g cm^{-1} s^{-1}}]$    \\ \hline
		$\rho_f$   &   $1$    &   $[\unit{g cm^{-3}}]$    \\ \hline
    $D_{\textrm{K}^+}$    &  $1.957 \cdot 10^{-5}$   & $[\unit{cm^{2} s^{-1}}]$ \\ \hline
    $D_{\textrm{Na}^+}$    &  $1.334 \cdot 10^{-5}$   & $[\unit{cm^{2} s^{-1}}]$ \\ \hline
    $D_{\textrm{Cl}^-}$    &  $2.033 \cdot 10^{-5}$   & $[\unit{cm^{2} s^{-1}}]$ \\ \hline
    $D_{\textrm{HCO}_3^-}$    &  $1.185 \cdot 10^{-5}$   & $[\unit{cm^{2} s^{-1}}]$ \\ \hline
    $D_{\textrm{Ca}^{++}}$    &  $7.92 \cdot 10^{-6}$   & $[\unit{cm^{2} s^{-1}}]$ \\ \hline
    $D_{\textrm{H}^+}$    &  $9.315 \cdot 10^{-5}$   & $[\unit{cm^{2} s^{-1}}]$ \\ \hline
	\end{tabular}
	\vspace*{4pt}
\caption{Values of model parameters.} 
\label{tab:parameters_model_KNa_pump}
\end{table}

\subsubsection{Electrical variables}

\begin{figure}[h!]
	\centering
	\begin{subfigure}[b]{0.495\textwidth}
		\includegraphics[width = \textwidth]{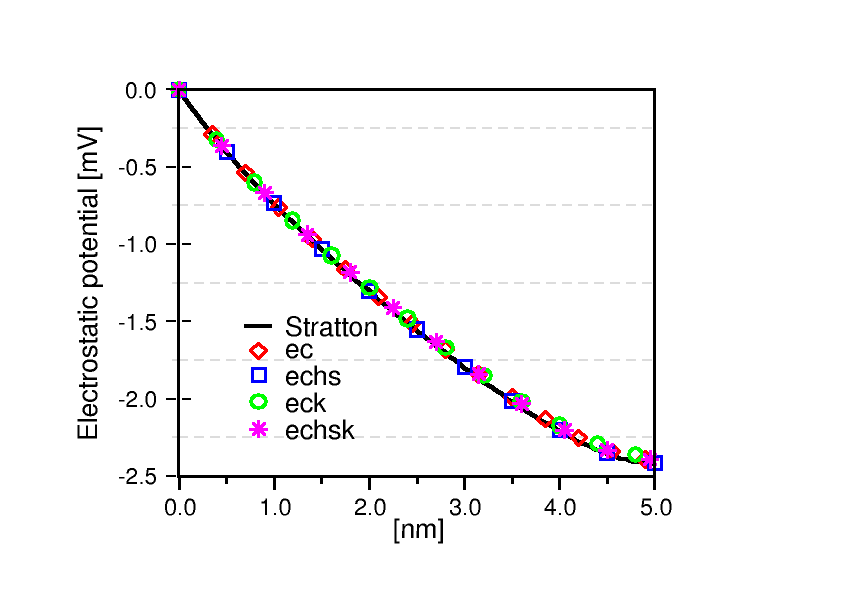}
		\caption{Electrostatic potential}
		\label{fig:pot_forzanti}	
	\end{subfigure}
	\begin{subfigure}[b]{0.495\textwidth}
		\includegraphics[width = \textwidth]{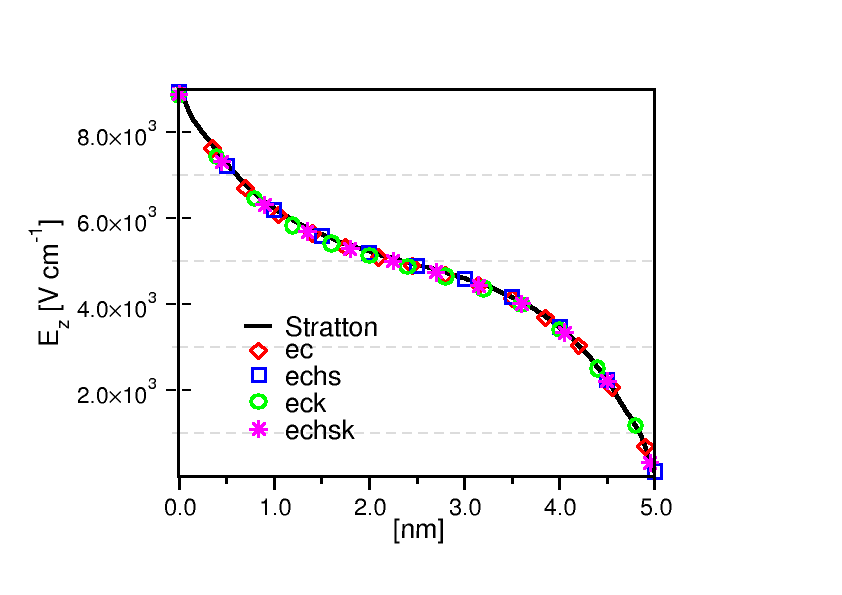}
		\caption{Electric field}
		\label{fig:Ez_forzanti}	
	\end{subfigure}
	\caption{Electrical variables along the axis of the channel.}
\end{figure}

\noindent
Figures~\ref{fig:Ez_forzanti} and~\ref{fig:pot_forzanti} show the spatial distributions 
of the electric potential and of the electric field inside the channel.
Since no permanent charge is included in the present simulation, the electric potential (and therefore also the electric field) 
is determined only by the Coulomb interaction among the ions in the channel.
This is the reason why electric field and electric potential distributions are scarcely affected by the different 
models of Section~\ref{sec:forzanti}.
Figure~\ref{fig:Ez_forzanti} also shows that the electric field profile is monotonic inside the channel.
This means, on one hand, that ions are transported with a constant direction depending on their sign 
(from the intracellular to the extracellular sides in the case of cations, from the extracellular to the intracellular 
sides in the case of anions), and, on the other hand, that ions cannot be trapped inside the channel, rather they are helped 
travel throughout the channel. 
The electrostatic potential shown in Figure~\ref{fig:pot_forzanti} allows us to perform a \textbf{\textit{first significant model 
comparison with experimental data}} {reported in} Table~\ref{tab:BioDataPot} where the measured value of the transepithelial 
membrane potential $V_m:= \varphi(0) - \varphi(L_{ch})$ is reported for various animals.
Results indicate that for all model choices of Section~\ref{sec:forzanti}, the simulated potential difference is 
in very good agreement with the data for monkeys~\cite{Chu1987} which can be considered as the animal species most similar to humans.

\begin{table}[h!]
	\centering
		\begin{tabular}{|l|l|l|}
    \hline
			$V_m \; [\unit{mV}] $ & Animal & Reference \\ \hline
			$3.80 \pm  0.26$ & ox & \cite{Cole61} \\ \hline
			$5.53 \pm 0.41$ & ox  & \cite{Cole62} \\ \hline
			$3.83 \pm 0.16$ & rabbit & \cite{Cole62} \\ \hline
			$-3.7\pm 0.3 $ & toad & \cite{Watanabe78} \\ \hline
			$-1.2 \pm 0.1$ & rabbit &  \cite{Krupin84} \\ \hline
			$-1.35 \pm 0.08$ & dog &  \cite{Iizuka1984} \\ \hline
			$\boxed{-2.5 \pm 0.2}$ & monkey	& \cite{Chu1987} \\ \hline
	\end{tabular}
	\vspace*{4pt}
	\caption{Experimental measurements for the transepithelial membrane potential. 
		The boxed value is the measured data for monkeys and is {considered} as the reference for comparison with our model simulations.} 
	\label{tab:BioDataPot}
\end{table}

\subsubsection{Chemical variables}

\begin{figure}[h!]
	\centering
	\begin{subfigure}[b]{0.495\textwidth}
		\includegraphics[width = \textwidth]{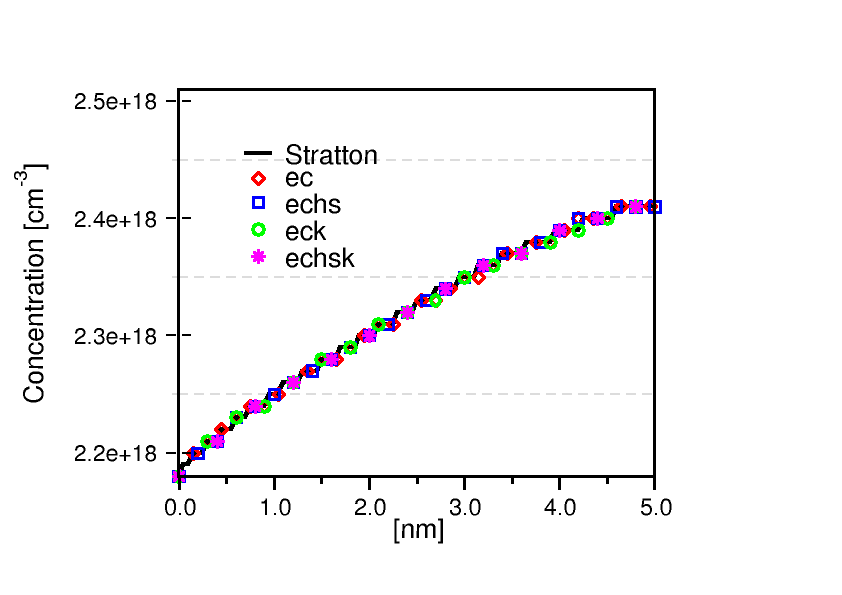}
		\caption{Spatial distribution of K$^+$.}
		\label{fig:conc_K}	
	\end{subfigure}
	\begin{subfigure}[b]{0.495\textwidth}
		\includegraphics[width = \textwidth]{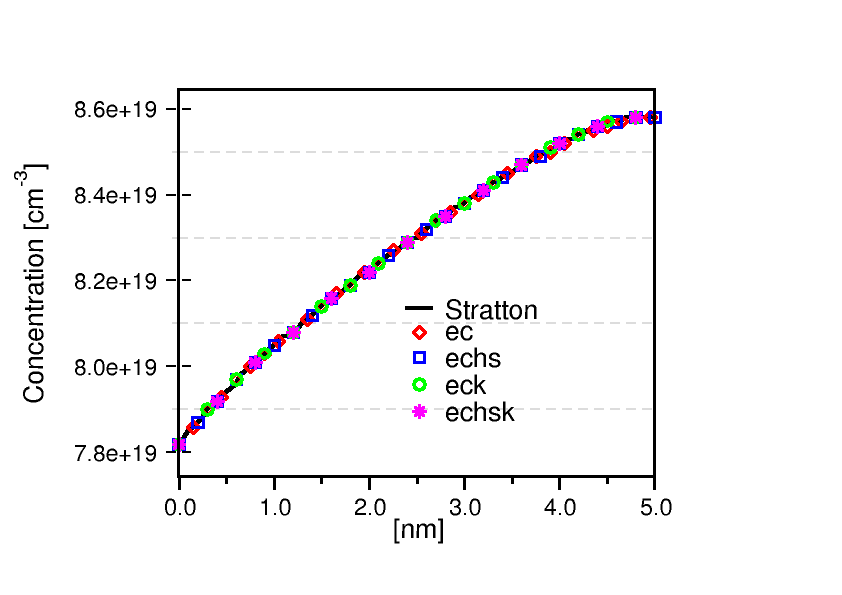}
		\caption{Spatial distribution of Na$^+$.}
		\label{fig:conc_Na}	
	\end{subfigure}
	
	\begin{subfigure}[b]{0.495\textwidth}
		\includegraphics[width = \textwidth]{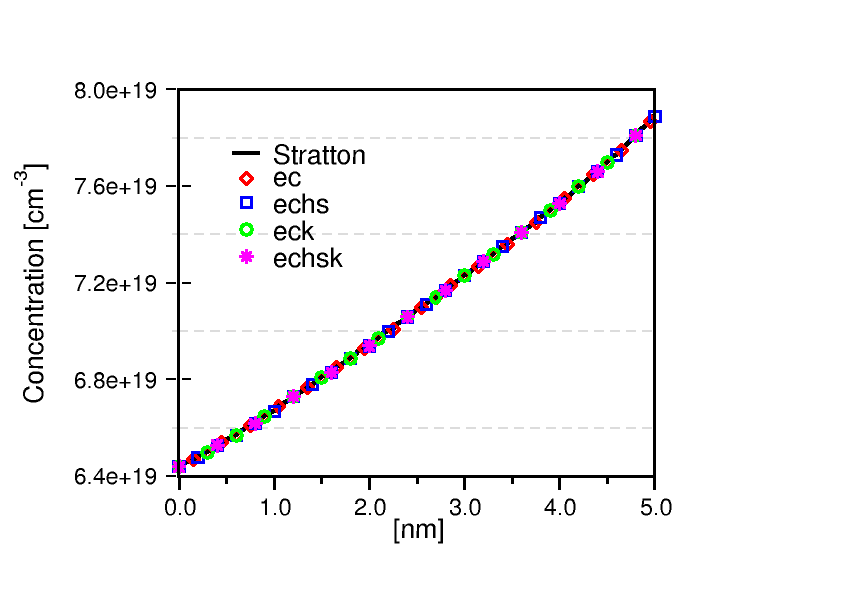}
		\caption{Spatial distribution of Cl$^-$.}
		\label{fig:conc_Cl}	
	\end{subfigure}
	\begin{subfigure}[b]{0.495\textwidth}
		\includegraphics[width = \textwidth]{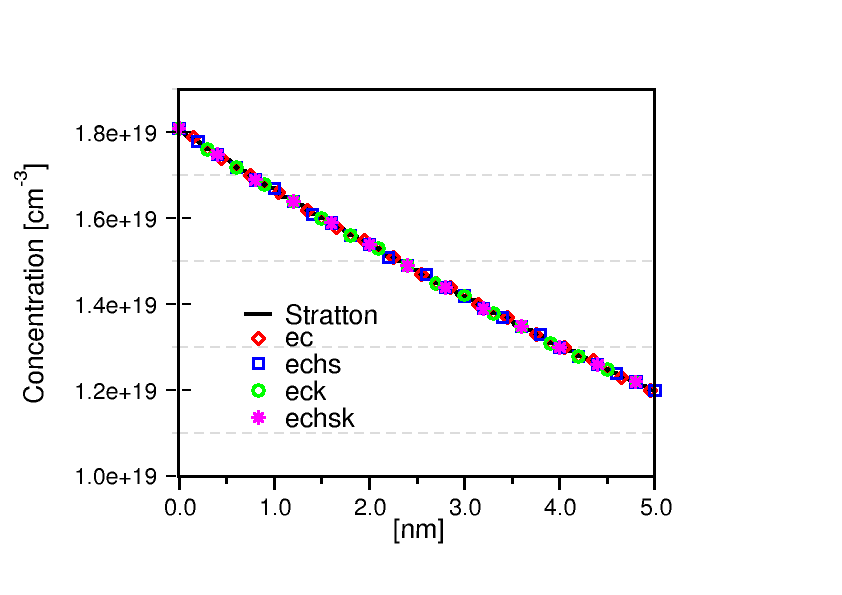}
		\caption{Spatial distribution of HCO$_3^-$.}
		\label{fig:conc_HCO3}	
	\end{subfigure}
	\caption{Ionic concentration spatial distribution}
\end{figure}

\noindent
Figures~\ref{fig:conc_K} and~\ref{fig:conc_Na} show the spatial distributions of cations inside the channel whereas 
Figures~\ref{fig:conc_Cl} and~\ref{fig:conc_HCO3} show the spatial distributions of anions inside the channel. 
Consistently with the simulated electric field and potential distributions, results indicate that the different models 
of the volume force density scarcely affect the ionic concentrations indicating that the difference
in fluid velocity, shown in the next section, are not strong enough to modify the ionic profiles.
As a second comment, we see that the spatial distribution of each ionic concentration is not linear inside the channel because 
of the presence of the electric field that is responsible for the drift contribution in the ion flux constitutive 
relation~\eqref{eq:PNP_flux}. 
The data reported in Table~\ref{tab:Current_Density_forzanti} allow us to perform a \textbf{\textit{second significant model comparison {with experimental data}}}. 
The data include the computed value of the axial component of the potassium current density at the extracellular side of the
channel $Z=L_{ch}$ and the computed value of the axial component of the sodium current density at the intracellular side 
of the channel $Z=0$ {for} the various choices for the model of the volume force density $\uF_{ion}$ 
illustrated in Section~\ref{sec:forzanti}. To allow a quantitative verification of the 
biophysical correctness of the predicted exchange of potassium and sodium ions across the channel, 
we introduce the following parameter
\begin{align}
& r:= \dfrac{g_{\textrm{K}^+}}{g_{\textrm{Na}^+}}, & \label{eq:ratio_stoichiometric}
\end{align}
{where $g_{\textrm{K}^+}$ denotes the boundary value of the flux density of potassium ions that 
enter into the {cell} and $g_{\textrm{Na}^+}$ denotes the boundary value of the 
flux density of sodium ions that flow out of the {cell}. According to the data of Table~\ref{tab:Ion+}, 
we have $r=2:3$.} The above parameter expresses the biophysical consistency of the 
boundary data {adopted in the numerical simulation because it} 
coincides with the {theoretically expected stoichiometric ratio} 
of the K$^+$ and Na$^+$ ions exchanged ($2:3$) by the sodium-potassium 
pump as represented in the schematic picture of Figure~\ref{KNapump}.
Because of the continuum approach employed in our model, {we are going to check 
the correct functionality of the simulated pump by computing the following parameter
\begin{align}
& \mathcal{R}:= \Big|\dfrac{J_{Z,\textrm{K}^+}}{J_{Z,\textrm{Na}^+}}\Big|, & \label{eq:ratio_stoichiometric_currents}
\end{align}
where $J_{Z,\textrm{K}^+}$ is the potassium ionic current density at the extracellular side of the channel
and $J_{Z,\textrm{Na}^+}$ is the sodium ionic current density at the intracellular side of the channel.}
The parameter $\mathcal{R}$ is the counterpart of the quantity $r$ defined in~\eqref{eq:ratio_stoichiometric}
and is quite sensitive to the choice of the volumetric force.
As a first comment, the results of Table~\ref{tab:Current_Density_forzanti} show that for each considered model of $\uF_{ion}$ 
the computed potassium current density is negative whereas the computed sodium current density is positive. 
This is consistent with the physiological function of the sodium-potassium pump because sodium ions flow out of the 
cell and potassium ions flow into the cell. 
As a second comment, the computed values of $\mathcal{R}$ indicate that agreement with the theoretical expected 
ratio 2:3 is achieved only in the case of the Stratton model and of the eck model, whereas the values 
of $\mathcal{R}$ computed with the other models are not in a feasible range. 
This allows us to conclude that the VE-PNP model predicts a correct direction of ion flow for 
the sodium-potassium pump in a good quantitative agreement with the stoichiometric ratio of the pump
only if the \textbf{Stratton} or the \textbf{eck} model are adopted to mathematically represent the volume force density in the 
linear momentum balance equation for the aqueous intracellular fluid.

\begin{table}[h!]
\begin{tabular}{{|l|l|l|l|}}
	\hline 
		Model for $\uF_{ion}$ & $J_{Z,\textrm{K}^+} [\unit{A cm^{-2}}]$ &  
		$J_{Z, \textrm{Na}^+} [\unit{A cm^{-2}}]$ & $\mathcal{R}$ \\ \hline \hline
		Stratton & $-0.064$ & $ 0.098 $ & $\boxed{0.653}$ \\ \hline
		ec & $-0.054$ & $0.44$ & $0.123$ \\ \hline
		echs & $-0.046 $ & $0.74$ & $0.062$ \\ \hline
		eck & $-0.064$ & $0.085$ & $\boxed{0.753}$ \\ \hline
		echsk & $-0.056$ & $0.38$ & $0.147$ \\ \hline 
	\end{tabular}
	\vspace*{4pt}
	\caption{Computed values of the axial component of the current density for sodium and potassium.
		The value $J_{Z,\textrm{K}^+}$ is computed at $Z=L_{ch}$ whereas the value $J_{Z,\textrm{Na}^+}$ is computed at $Z=0$. 
		We set $\mathcal{R}:= |J_{Z,\textrm{K}^+}/J_{Z,\textrm{Na}^+}|$. 
		The boxed values indicate the best model predictions to be compared with the theoretical expected ratio 2:3.}
	\label{tab:Current_Density_forzanti}
\end{table}

\subsubsection{Fluid variables}

\begin{figure}[h!]
	\centering
	\begin{subfigure}[b]{0.55\textwidth}
		\includegraphics[width = \textwidth]{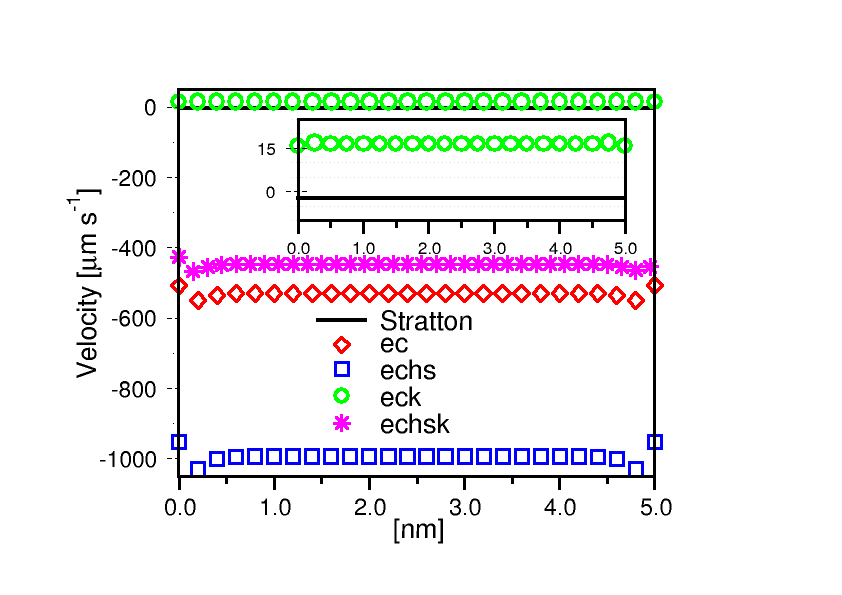}
		\caption{Aqueous humor fluid velocity}
		\label{fig:vel_forzanti}	
	\end{subfigure}
	\begin{subfigure}[b]{0.44\textwidth}
		\includegraphics[width = \textwidth]{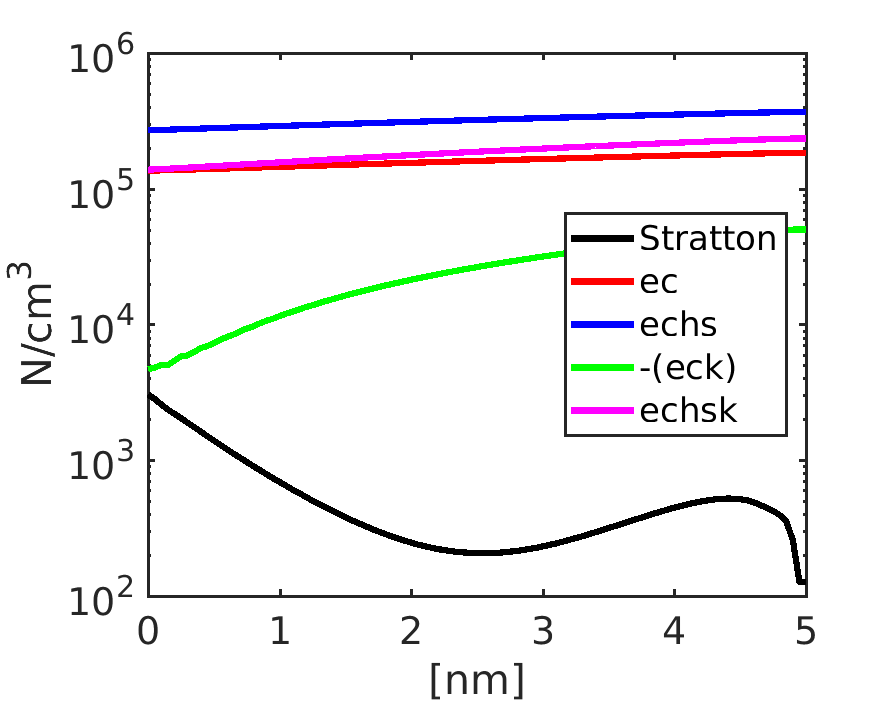}
		\caption{Volumetric force density}
		\label{fig:force_calc_forzanti}	
	\end{subfigure}
	\caption{Fluid variables along the axis of the channel.}
\end{figure}
Figures~\ref{fig:vel_forzanti} and~\ref{fig:force_calc_forzanti} show the spatial distributions of the component
of the fluid velocity and of the volumetric force density along the $Z$ axis.
{Predicted velocities} are all negative 
except {in the case of the eck model}; {similarly,} the 
{computed} volumetric force densities are all positive
except {in} the case of the eck model 
(in Fig.~\ref{fig:force_calc_forzanti} we have reported the absolute value of the force density). 
{Moreover it is easy to check that} 
the variation of the velocity is one-to-one correlated with the
variation of the volumetric force density because of the homogeneous initial and boundary conditions 
that are applied to the Stokes system. 

\begin{table}[h!]
	\centering
	\begin{tabular}{|l|l|}
		\hline 
		Model for $\uF_{ion}$ & $\bar{v}_Z [\unit{\mu m \; s^{-1}}]$ \\ \hline \hline
		Stratton & $-2.34$ \\ \hline
		ec & $-529.5$  \\ \hline
		echs &  $-992$ \\ \hline
		eck & $\boxed{16.54}$ \\ \hline 
		echsk & $-445.6$ \\  \hline
	\end{tabular}
	\vspace*{4pt}
	\caption{Computed mean values of the axial component of the intrachannel water fluid velocity. 
		The boxed value indicates the sole model results that are in agreement with an outflux of
		aqueous humor from the cell into the extracellular side.}
	\label{tab:fluid_vel}
\end{table}

\noindent
{Table~\ref{tab:fluid_vel} reports the values of the fluid velocity computed at the center of the channel
for each model considered in Section~\ref{sec:forzanti}.}
Results allow us to perform a \textbf{\textit{third significant model comparison {with experimental data}}}: 
only by describing the volume force 
density through the \textbf{eck} model the VE-PNP formulation is able to predict
a positive fluid velocity which corresponds to the production of AH from the cell into the basolateral space.
More specifically, if we assume a value of $2.5 [\unit{\mu l \, s^{-1}}]$ for a normal AH flow through the eye pupil 
(cf.~\cite{Moses1987}) and an equivalent radius of $1 [\unit{mm}]$ for the eye pupil of an adult, we see that 
a physiological value of $v_Z$ is of about $14 [\unit{\mu m \, s^{-1}}]$ which agrees well with the value 
of $16.54 [\unit{\mu m \, s^{-1}}]$ predicted by the eck model. 
The other results from Table~\ref{tab:fluid_vel} (negative velocities) indicate that the magnitude of the
predicted AH flow is nonphysically large, except in the case of the {Stratton model, thus justifying its 
wide adoption in the literature.}

\subsection{Conclusions}
{The study of the interaction between the ionic component (Na$^+$-K$^+$ pump) and AH production through 
a mathematical continuum approach based on the VE-PNP model, under the condition of 
adopting the \textbf{eck} formulation of the volume force density that constitutes the source 
term in the fluid momentum balance equation, shows that simulation results are in agreement with:}
\begin{enumerate}
\item the experimentally measured value of transepithelial membrane potential;
\item the physiological stoichiometric rate of 2:3 that characterizes the sodium-potassium pump;
\item the direction of current densities of sodium (flowing out of the cell) and potassium (flowing into the cell);
\item the direction of AH  flow (outward the cell);
\item the magnitude of AH fluid velocity.
\end{enumerate}

\noindent
The aferomentioned results {support} the mathematical and biophysical 
{motivation to adopt the \textbf{eck} model
in the remainder of the article where we introduce a more realistic geometrical description of the 
ionic channel and we include the main ionic pumps 
to describe the electric pressure exerted by the ions on the intrachannel fluid.}

\section{Cellular scale simulation of ionic pumps in AH production}\label{sec:simulations_2}
In this section we {use the VE-PNP model to
carry out an extensive quantitative investigation on the active role of the ionic exchanges that
are identified in~\cite{Kiel2001} and~\cite{Kiel2011} as important determinants in AH secretion.
To this purpose,} we adopt the VE-PNP formulation in which 
the volumetric force exerted from ions onto the fluid is described {by} the eck model 
{illustrated in Section~\ref{ch:eck}. In addition, we employ in the numerical
simulations a more realistic {ionic} channel geometry than {that shown} in Figure~\ref{fig:cilinder_r04FITTA}, 
obtained by including in the computational domain a small amount of cell membrane 
as well as the presence of the antichambers.}
The {unified modeling and computational framework is here applied to study
the function of each ionic pump illustrated in Section~\ref{sec:AH} with the goal
of examining the output results, such as electrostatic potential, stoichiometric ratios 
and AH velocity, as functions of 
the input parameters, such as the osmotic coefficient, the value of the permanent 
electric charge density and the non-homogeneous Neumann boundary conditions for ion flux densities.}

\begin{figure}[h!]
	\centering
	\includegraphics[width = 0.65\textwidth]{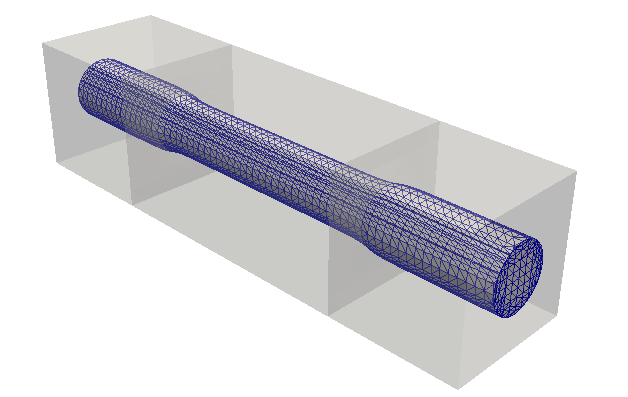}
	\caption{Computational domain for the simulation of the ionic pumps and channels involved in AH production. 
		The two external cylinders represent the channel antichambers whereas the central cylinder is the channel region. 
		The partition of the channel domain into about 110391 tetrahedral elements is illustrated,
		whereas the mesh partition of the region surrounding the channel is not shown for visual clarity.}
	\label{fig:mech}	
\end{figure}

\noindent
Figure~\ref{fig:mech} shows the geometrical structure constituting the computational domain. 
The parallelepiped containing the cylinder represents the regions in which the transmembrane channel is divided, {namely}
two external cylinders {representing} the channel antichambers {and a} central cylinder {representing the} ionic channel.
The parallelepiped is composed by the union of two cubes of side equal to $2.5 [\unit{nm}]$ and by a central parallelepiped 
of length equal to $5 [\unit{nm}]$ in such a way that the total length is equal to $10 [\unit{nm}]$.
The portion of the cylindrical structure inside the two external cubes has a radius of $0.6 [\unit{nm}]$ whereas 
the portion inside the central parallelepiped has a radius of $0.4 [\unit{nm}]$. 
The adopted geometrical representation is based on the biophysical setting 
analyzed in~\cite{Flux_coupling,one_conformation} and aims at reproducing the morphology 
of a realistic protein membrane channel, where the two external cylinders play the role of channel 
antichambers and the cylinder at the center plays the role of the channel region in which the main 
electrochemical and fluid processes take place.
The full domain has been partitioned in tetrahedral as reported in Figure~\ref{fig:mech} where 
the discretization of the parallelepiped
surrounding the cylinder is not shown for sake of visual clarity.
{Referring to the notation of Figure~\ref{fig:bordo_mech}, in the remainder of the section,}
the VE-PNP equations~\eqref{eq:PNP}-~\eqref{eq:PNP_flux} and the Stokes system
~\eqref{eq:stokes_incomp}-~\eqref{eq:strain_rate} {are solved}
only inside the cylinder $\Omega_{INT}$ whereas the Poisson 
equation~\eqref{eq:PNP_Poisson}-~\eqref{eq:PNP_potential}) {is solved} in the whole domain $\Omega$.

\subsection{Boundary and initial conditions}\label{sec:bcs_ics_2}
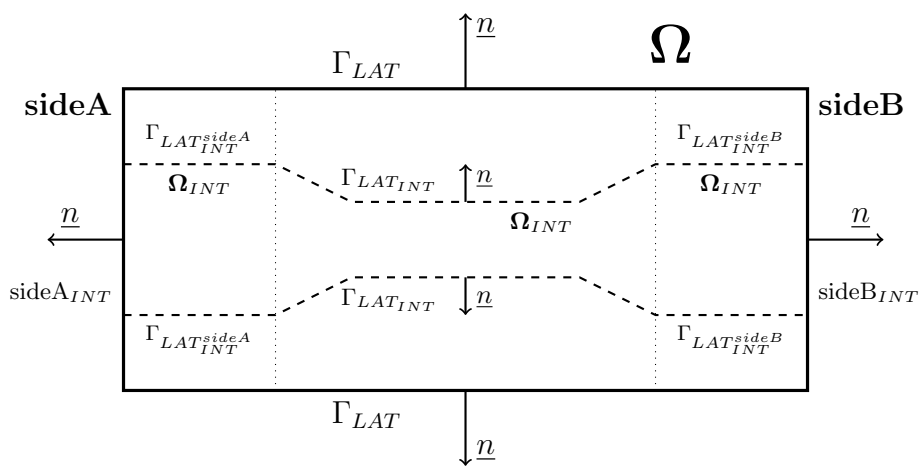
\begin{figure}[h!]
	\centering
	\begin{tikzpicture}
	\draw[very thick] (-4,-2) -- (5,-2)
		-- (5,2) -- (-4,2) -- cycle;
	\draw[dashed, thick] (-4,-1) -- (-2,-1)
		-- (-1,-0.5) -- (2,-0.5) -- (3,-1) -- (5,-1);
	\draw[dashed, thick] (-4,1) -- (-2,1)
		-- (-1,0.5) -- (2,0.5) -- (3,1) -- (5,1);	
	\draw[dotted] (-2,-2) -- (-2,2);
	\draw[dotted] (3,-2) -- (3,2);
	\draw[thick,->] (-4,0) -- (-5,0); \node[above] at (-4.7,0) {$\underline{n}$};
	\draw[thick,->] (5,0) -- (6,0); \node[above] at (5.7,0) {$\underline{n}$};
	\draw[thick,->] (0.5,2) -- (0.5,3); \node[right] at (0.5,2.8) {$\underline{n}$};
	\draw[thick,->] (0.5,-2) -- (0.5,-3); \node[right] at (0.5,-2.8) {$\underline{n}$};
	\draw[thick,<-] (0.5,1) -- (0.5,0.5); \node[right] at (0.5,0.8) {\footnotesize $\underline{n}$};
	\draw[thick,<-] (0.5,-1) -- (0.5,-0.5); \node[right] at (0.5,-0.8) {\footnotesize $\underline{n}$};
	\node[above] at (3.2,2.2) {\LARGE $\boldsymbol{\Omega}$};
	\node[below] at (-3,1) {\scriptsize$\boldsymbol{\Omega}_{INT}$};
	\node[below] at (4,1) {\scriptsize$\boldsymbol{\Omega}_{INT}$};
	\node[below] at (1.5,0.5) {\scriptsize$\boldsymbol{\Omega}_{INT}$};
	\node[left] at (-4,1.8) {\textbf{sideA}};	
	\node[left] at (-4,-0.7) {\scriptsize{sideA$_{INT}$}};
	\node[right] at (5,1.8) {\textbf{sideB}};	
	\node[right] at (5,-0.7) {\scriptsize{sideB$_{INT}$}};	
	\node[above] at (-0.8,2) {$\Gamma_{LAT}$};
	\node[below] at (-0.8,-2) {$\Gamma_{LAT}$};
	\node[above] at (-0.5,0.5) {\scriptsize{$\Gamma_{LAT_{INT}}$}};	
	\node[below] at (-0.5,-0.5) {\scriptsize{$\Gamma_{LAT_{INT}}$}};
	\node[above] at (-3,1) {\tiny{$\Gamma_{LAT_{INT}^{sideA}}$}};		
	\node[below] at (-3,-1) {\tiny{$\Gamma_{LAT_{INT}^{sideA}}$}};
	\node[above] at (4,1) {\tiny{$\Gamma_{LAT_{INT}^{sideB}}$}};	
	\node[below] at (4,-1) {\tiny{$\Gamma_{LAT_{INT}^{sideB}}$}};
\end{tikzpicture}
	\caption{Two-dimensional cross-section of the channel geometry and boundary labels for the simulation of AH production.}
	\label{fig:bordo_mech}
\end{figure}

In Section~\ref{sec:simulations_1} we have highlighted the fundamental role played by the BCs 
in the simulation of the sodium-potassium pump. Because here we are treating a {wider variety
of ion exchangers, a more complex computational domain as well as the presence of a larger number of ion
species, to help the clarity of the discussion we report in 
Figure~\ref{fig:bordo_mech}} a two-dimensional cross-section of the channel geometry and 
{identify} the various regions of the domain with the corresponding labels for further reference.

\subsubsection{Poisson equation}
Because of the presence of the interface between the internal cylinder
and the surrounding parallelepiped we need to treat the jump of the electric displacement
across that interface as well as the possible presence of electrical charge on this interface.
Let $\gamma:= \Gamma_{LAT_{INT}^{sideA}} \cup \Gamma_{LAT_{INT}^{sideB}} \cup \Gamma_{LAT_{INT}}$ denote 
the two-dimensional surface separating the channel region from the surrounding membrane region as depicted in 
Figure~\ref{fig:bordo_mech}.
For a given vector-valued function $\underline{\tau}: \Omega \rightarrow \mathbb{R}^3$ we define the jump 
of $\underline{\tau}$ across the surface $\gamma$ as 
$$
\jmp{\underline{\tau}}_\gamma:= \left(\underline{\tau}|_{\Omega \setminus
\Omega_{INT}}|_{\gamma} - \underline{\tau}|_{\Omega_{INT}}|_{\gamma} \right) \cdot \un,
$$
whereas for a given scalar-valued function $\phi:\Omega \rightarrow \mathbb{R}$ we define the jump of $\phi$ across 
the surface $\gamma$ as
$$
\jmp{\phi}_\gamma:= \phi|_{\Omega \setminus \Omega_{INT}}|_{\gamma} \un
- \phi|_{\Omega_{INT}}|_{\gamma} \un.
$$

\noindent
We notice that the jump of a vector-valued function is a scalar function whereas the jump of a scalar-valued 
function is a vector function.
For all $t \in I_T$, the BCs for the Poisson equation~\eqref{eq:PNP_Poisson}-\eqref{eq:PNP_potential} are:

\begin{subequations}\label{eq:BC_s_Poisson_2}
\begin{align}
& \varphi = 0                     & \qquad \textrm{on } sideA_{int} & \\
& \uD \cdot\un = 0 & \qquad \textrm{on } \Gamma_{LAT} \cup sideB \cup sideA \cup sideB_{int} & 
\label{eq:Neumann_homog_bc_outlet} \\
& \jmp{\uD}_\gamma = h_\gamma & \qquad \textrm{on } \gamma & \label{eq:jump_D} \\
& \jmp{\varphi}_\gamma = 0 & \qquad \textrm{on } \gamma, & \label{eq:jump_phi}
\end{align}
\end{subequations}
where:
$$
h_\gamma = 
\left\{
\begin{array}{ll}
\sigma_{fixed} & \qquad \textrm{on } \Gamma_{LAT_{INT}} \\
0              & \qquad \textrm{on } \Gamma_{LAT_{INT}^{sideA}} \cup \Gamma_{LAT_{INT}^{sideB}}, 
\end{array}
\right.
$$ 

\noindent
$\sigma_{fixed}$ $[\unit{C \; m^{-2}}]$ being a given distribution of superficial permanent charge density 
that mathematically represents the electric charge contained in the aminoacid structure of the protein surrounding the ion channel.
We notice that the interface condition~\eqref{eq:jump_phi} expresses the physical fact that the electric potential is a 
continuous function across $\gamma$, whereas the interface condition~\eqref{eq:jump_D} expresses the physical fact that 
the normal component of the displacement vector is discontinuous across the surface separating the channel region and the 
lipid membrane bilayer because of the presence of aminoacid fixed charge density $\sigma_{fixed}$.
The value of $\sigma_{fixed}$ {needs be determined in order to
reproduce} the correct functionality of the several ionic pumps/exchangers.
To this purpose, a simulation campaign has to be performed to heuristically tune-up the 
values of $\sigma_{fixed}$ (see~\cite{Art_equations} for the sodium-potassium pump). 
The results of this procedure in the present context are reported in Table~\ref{tab:sigma_fixed}. 
\begin{table}[h!]
\begin{tabular}{{|l|c|}}
	\hline 
		Pump/channel &  $\sigma_{fixed} [\unit{C \; cm^{-2}}]$ \\ \hline \hline
		sodium-potassium pump      &  $ -1\cdot 10^{10}$  \\ \hline
		calcium-sodium pump        &  $-1.2\cdot 10^{12}$ \\ \hline
		anion channel         &  $+3.9\cdot 10^{11}$ \\ \hline
		hydrogenate-sodium pump   &  $-2.65\cdot 10^{12}$ \\ \hline
	\end{tabular}
	\vspace*{4pt}
	\caption{Values of the fixed charge density $\sigma_{fixed}$.}
	\label{tab:sigma_fixed}
\end{table}

\subsubsection{Nernst-Planck equations}
The several ionic pumps/exchangers involved in  AH production are simulated by considering the contribution
of different ions in order to produce the correct electrostatic potential drop across the
cell membrane. The list of these ions is reported below.
In {complete analogy with what done in Section~\ref{sec:simulations_1} for the BCs and ICs of
the Nernst-Planck equations~\eqref{eq:PNP}-~\eqref{eq:PNP_flux},} we report 
in Tables~\ref{tab:bc_ic_NaKpump}-~\ref{tab:bc_ic_Hydro_sodium_pump} 
the BCs and ICs {adopted to reproduce the correct biophysical functionality
of each ion pump.}

\begin{description}
 \item [Na$^+$-K$^+$ pump]: Na$^+$, K$^+$, Cl$^-$, HCO$_3^-$ are treated.
 \item [Ca$^{++}$-Na$^+$ pump]: Na$^+$, K$^+$, Cl$^-$, HCO$_3^-$, Ca$^{++}$ are treated.
 \item [Cl$^-$-HCO$_3^-$ pump]: Na$^+$, K$^+$, Cl$^-$, HCO$_3^-$ are treated.
 \item [H$^+$-Na$^+$ pump]: Na$^+$, K$^+$, Cl$^-$, HCO$_3^-$, H$^+$ are treated.
\end{description}

\begin{table}[h!]
		\begin{tabular}{|l|l|l|}
		\hline
			Na$^+$ = Na$^+_{in}$ & $\uf_{\textrm{K}^+}\cdot\un = 
			g_{\textrm{K}^+}$ & \textrm{on} $sideA_{int} $ 
			\\ \hline
			K$^+$ = K$^+_{out}$ & $\uf_{\textrm{Na}^+}\cdot\un = 
			g_{\textrm{Na}^+}$ & \textrm{on} $sideB_{int} $ 
			\\ \hline
			$\uf_{\textrm{Na}^+} \cdot\un = 0$ & $\uf_{\textrm{K}^+}\cdot\un = 0 $ 
			& \textrm{on} $\gamma$ \\ \hline
			K$^{+^0}(x) = $ K$^+_0$ & Na$^{+^0}(x) = $ Na$^+_0$ & \textrm{in} $\Omega_{INT}$ \\	\hline
			Cl$^- = $ Cl$^-_{in}$ & HCO$_3^- = $ HCO$^-_{3_{in}}$ & \textrm{on} $sideA_{int}$\\ \hline
			Cl$^- = $ Cl$^-_{out}$ & HCO$_3^- = $ HCO$^-_{3_{out}}$ &  \textrm{on} $sideB_{int} $\\ \hline
			$\uf_{\textrm{Cl}^-}\cdot\un = 0$ & 
			$\uf_{\textrm{HCO}_3^-}\cdot\un = 0 $ & 
			\textrm{on} $\gamma$  
			\\ \hline
			Cl$^{-^0}(x) = $ Cl$^-_0$ & HCO$_3^{-^0}(x) = $ HCO$_{3_0}^-$ & \textrm{in} $\Omega_{INT}$ 
			\\ \hline
	\end{tabular}
	\vspace*{4pt}
	\caption{BCs and ICs for the sodium-potassium pump.}
	\label{tab:bc_ic_NaKpump}		
\end{table}

\begin{table}[h!]
		\begin{tabular}{|l|l|l|l|}
		\hline
		K$^+ = $ K$^+_{in}$ & Na$^+ = $ Na$^+_{in}$ & $\uf_{\textrm{Ca}^{++}}\cdot\un = 
		g_{\textrm{Ca}^{++}} $ & \textrm{on} $sideA_{int}$ \\ \hline
		K$^+ = $ K$^+_{out}$ & Ca$^{++} = $ Ca$^{++}_{out}$ & $\uf_{\textrm{Na}^+}\cdot\un = 
		g_{\textrm{Na}^+}$ 
		& \textrm{on} $sideB_{int}$ \\ \hline
		$\uf_{\textrm{Ca}^{++}}\cdot\un = 0$ & $\uf_{\textrm{Na}^+}\cdot\un = 0 $ & 
		$\uf_{\textrm{K}^+}\cdot\un = 0 $ 
		& \textrm{on} $\gamma$  \\ \hline
		K$^{+^0}(x) = $ K$^+_0$ & Na$^{+^0}(x) = $ Na$^+_0 $ & Ca$^{{++}^0}(x) = $ Ca$^{++}_0 $ 
		& \textrm{in} $ \Omega_{INT} $\\ \hline
		Cl$^- = $ Cl$^-_{in}$ & HCO$_3^- = $ HCO$^-_{3_{in}} $ & & \textrm{on} $ sideA_{int} $\\ \hline
		Cl$^- = $ Cl$^-_{out}$ & HCO$_3^- = $ HCO$^-_{3_{out}} $ & & \textrm{on} $ sideB_{int} $\\ \hline
		$\uf_{\textrm{Cl}^-}\cdot\un = 0 $ & $\uf_{\textrm{HCO}_3^-}\cdot\un = 0 $ & & 
		\textrm{on} $\gamma$ \\ \hline
		Cl$^{-^0}(x) = $ Cl$^-_0 $ & HCO$_3^{-^0}(x) = $ HCO$_{3_0}^-$ & & \textrm{in} $ \Omega_{INT} $ 
		\\ \hline
\end{tabular}
\vspace*{4pt}
 \caption{BCs and ICs for the calcium-sodium pump.}
 \label{tab:bc_ic_CaNapump}		
\end{table}

\begin{table}[h!]
		\begin{tabular}{|l|l|l|}
		\hline
			K$^+ = $ K$^+_{in} $ & Na$^+ = $ Na$^+_{in} $ & \textrm{on} $ sideA_{int} $ \\ \hline
			K$^+ = $ K$^+_{out}$ & Na$^+ = $ Na$^+_{out} $ & \textrm{on} $ sideB_{int} $\\ \hline
			$\uf_{\textrm{Na}^+}\cdot\un = 0$ & $\uf_{\textrm{K}^+}\cdot\un = 0 $ & 
			\textrm{on} $\gamma$ 	\\ \hline
			K$^{+^0}(x) = $ K$^+_0 $ & Na$^{+^0}(x) = $ Na$^+_0 $ & \textrm{in} $\Omega_{INT} $ \\ \hline
			$\uf_{\textrm{Cl}^-}\cdot\un = g_{\textrm{Cl}^-} $ & 
			$\uf_{\textrm{HCO}_3^-}\cdot\un = g_{\textrm{HCO}_3^-} $ & 
			\textrm{on} $ sideA_{int} $\\ \hline
			Cl$^- = $ Cl$^-_{out} $ & HCO$_3^- = $ HCO$^-_{3_{out}} $ & \textrm{on} $ sideB_{int} $\\ \hline
			$\uf_{\textrm{Cl}^-}\cdot\un = 0 $ & $\uf_{\textrm{HCO}_3^-}\cdot\un = 0 $ & 
			\textrm{on} $\gamma$ \\ \hline
			Cl$^{-^0}(x) = $ Cl$^-_0 $ & HCO$_3^{-^0}(x) = $ HCO$_{3_0}^- $ & \textrm{in} $ \Omega_{INT}	$ 
			\\ \hline
\end{tabular}
\vspace*{4pt}
 \caption{BCs and ICs for the anion channel.}
 \label{tab:bc_ic_Anion_channel}		
\end{table}
\begin{table}[h!]
		\begin{tabular}{|l|l|l|l|}
		\hline
		K$^+ = $ K$^+_{in} $ & Na$^+ = $ Na$^+_{in}$ & $\uf_{\textrm{H}^+}\cdot\un = 
		g_{\textrm{H}^+} $ & 
		\textrm{on} $ sideA_{int}$ \\ \hline
		K$^+ = $ K$^+_{out}$ & H$^+ = $ H$^+_{out}$ & $\uf_{\textrm{Na}^+}\cdot\un = 
		g_{\textrm{Na}^+} $ & 
		\textrm{on} $ sideB_{int} $\\ \hline
		$\uf_{\textrm{H}^+}\cdot\un = 0$ & $\uf_{\textrm{Na}^+}\cdot\un = 0$ & 
		$\uf_{\textrm{K}^+}\cdot\un = 0 $ 
		& \textrm{on} $\gamma$ \\ \hline
		K$^{+^0}(x) = $ K$^+_0 $ & Na$^{+^0}(x) = $ Na$^+_0 $ & H$^{+^0}(x) = $ H$^{+}_0 $ & 
		\textrm{in} $ \Omega_{INT} $\\ \hline
		Cl$^- = $ Cl$^-_{in}$ & HCO$_3^- = $ HCO$^-_{3_{in}}$ & & \textrm{on} $ sideA_{int} $\\ \hline
		Cl$^- = $ Cl$^-_{out}$ & HCO$_3^- = $ HCO$^-_{3_{out}}$ & & \textrm{on} $ sideB_{int} $\\ \hline
		$\uf_{\textrm{Cl}^-}\cdot\un = 0 $ & 
		$\uf_{\textrm{HCO}_3^-}\cdot\un = 0 $ & & \textrm{on}
		$\gamma$ \\ \hline
		Cl$^{-^0}(x) = $ Cl$^-_0$ & HCO$_3^{-^0}(x) = $ HCO$_{3_0}^-$ & & \textrm{in} $ \Omega_{INT} $ 
		\\ \hline
\end{tabular}
\vspace*{4pt}
 \caption{BCs and ICs for the hydrogenate-sodium pump.}
 \label{tab:bc_ic_Hydro_sodium_pump}		
\end{table}

\noindent
The values of the boundary data for the ionic pumps and ionic channels are specified 
in Tables~\ref{tab:sp_pump_2}-~\ref{tab:hyd_s_pump_2}.

\subsubsection{Stokes system}
{We adopt the same BCs and ICs as in Section~\ref{sec:simulations_1}:}, 
\begin{subequations}\label{eq:BC_s_Stokes_2}
\begin{align}
& \uu = \underline{0}                     & \qquad \textrm{on } \gamma & \\
& \underline{\underline{{\sigma}}} \:\un = \underline{0} & \qquad \textrm{on } 
sideA_{int} \cup sideB_{int} &\\
& \uu(\ux,0) = \underline{0}              & \qquad \forall \ux \in \Omega_{INT}.&
\end{align}
\end{subequations}

As already pointed out, to describe the volumetric force at the right-hand side of the linear momentum balance equation in the 
Stokes system we use the {\textbf{eck}} model. 
The value of $k$ is considered a {characteristic} property of the single pump and channel and 
is reported in Table~\ref{tab:parameter_electro_osmotic}.

\begin{table}[h!]
\begin{tabular}{{|l|c|}}
	\hline 
		Pump/channel &  $k [\unit{N \; m}]$ \\ \hline 
		sodium-potassium pump      &  $4.1\cdot 10^{-19}$ \\ 
		calcium-sodium pump       &  $24\cdot 10^{-19}$ \\ 
		anion channel         &  $4.1\cdot 10^{-19}$\\
		hydrogenate-sodium pump    &  $4 \cdot 10^{-19}$\\ \hline
	\end{tabular}
	\vspace*{4pt}
	\caption{Values of the electrochemical osmotic parameter $k$ for each pump/channel involved in 
	the process of AH production.}
	\label{tab:parameter_electro_osmotic}
\end{table}

\subsection{Simulation results} 
Reported data for the vector-valued variables (such as electric field, current densities and velocity) are the $Z$ 
component of the vectors because the other two computed components were comparably negligible. 
We set $t_0=0 \; [\unit{s}]$ and $T_{obs} = 50 \; [\unit{ns}]$, 
a sufficiently large value to ensure that the simulated system has reached steady-state conditions at $t=T_{obs}$:
all the figures in the remainder of the section illustrate computed results at this time.
The values of the dielectric permittivity of the intrachannel fluid $\epsilon_r^{f}$, of the fluid shear viscosity 
$\mu_f$, of the fluid mass density $\rho_f$ and of the diffusion coefficients $D_i$ of each $i$-th ionic species 
involved in the computational tests are reported in Table~\ref{tab:parameters_model_KNa_pump}.

\subsubsection{Electrical variables}
\begin{figure}[h!]
	\centering
	\begin{subfigure}[b]{0.495\textwidth}
		\includegraphics[width = \textwidth]{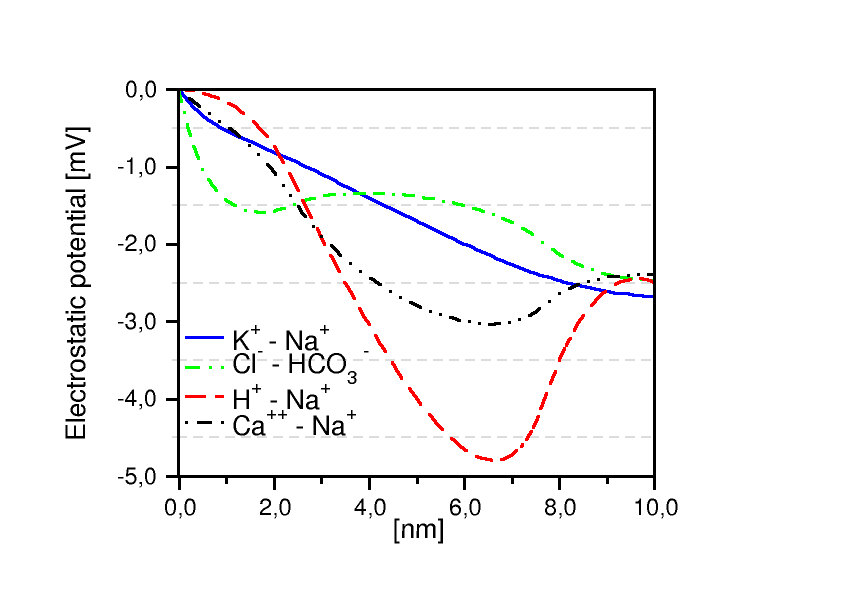}
		\caption{Electrostatic potential}
		\label{fig:mech_potential}	
	\end{subfigure}
	\begin{subfigure}[b]{0.495\textwidth}
		\includegraphics[width = \textwidth]{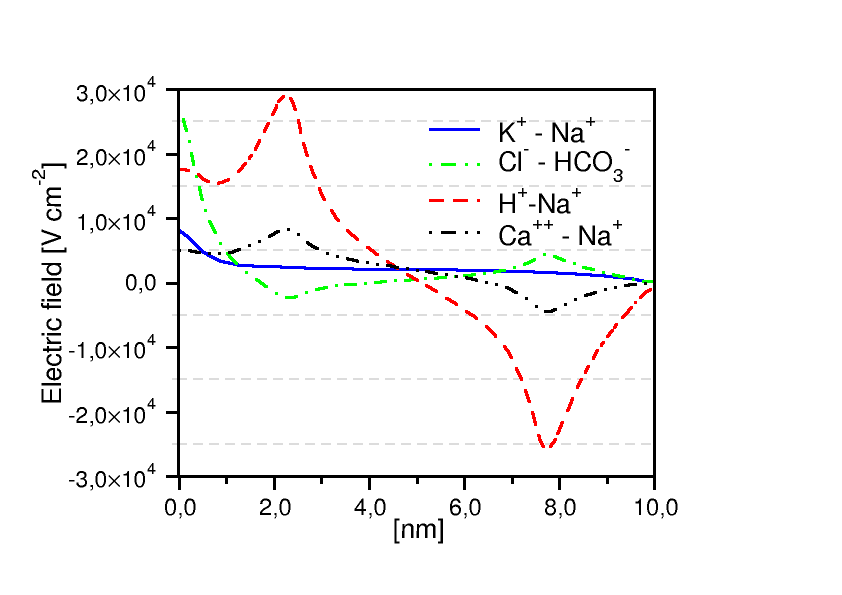}
		\caption{Electric field}
		\label{fig:mech_Efield}	
	\end{subfigure}
	\caption{Electrical variables along the axis of the channel.}
\end{figure}

Figure~\ref{fig:mech_potential} shows the transepithelial electrostatic potential as calculated by the simulations.
We note how the electric potential is strongly influenced by the presence of the fixed surface charge density $\sigma_{fixed}$ in the central region of the domain (cf.~Table~\ref{tab:sigma_fixed}),
{with particular emphasis in the case of the hydrogenate-sodium pump.}
It is remarkable to notice that, as in the case of the simulation of the sodium-potassium pump 
{illustrated} in Section~\ref{sec:simulations_1}, 
also in this more complex biophysical setting, for each simulated pump/channel, 
the computed value of \textbf{\textit{the transepithelial membrane potential is in very good agreement with the 
experimental data}} for monkeys reported in Table~\ref{tab:BioDataPot}.

Figure~\ref{fig:mech_Efield} shows the computed spatial behavior of the axial component $E_Z$ of the electric field
for each considered pump/channel.
Consistently with electrostatic potential, we see that in all simulations $E_Z$ is a monotonically function of position 
in the central region of the channel. Then, coming closer to the outlet section at $Z=L_{ch}$, all the simulated profiles
become flat in accordance with 
the homogeneous Neumann boundary condition~\eqref{eq:Neumann_homog_bc_outlet}.
Specifically, in the case of cation pumps, the electric field is decreasing along the central part of the channel
whereas in the case of the anion channel the electric field is increasing. 
These two opposite behaviors are related to the presence of surface charge on $\Gamma_{LAT_{IN}}$ of opposite 
sign (negative for cation pumps, positive for the anion pump). 
In the case of the hydrogenate-sodium pump, the electric field profile experiences a large increase in magnitude moving 
along the channel axis from the intracellular side towards the extracellular side because of the elevated 
negative fixed charge density distributed on the lateral surface on the channel region (cf.~Table~\ref{tab:sigma_fixed}). 

\subsubsection{Chemical variables}
\begin{figure}[h!]
	\centering
	\begin{subfigure}[b]{0.495\textwidth}
		\includegraphics[width = \textwidth]{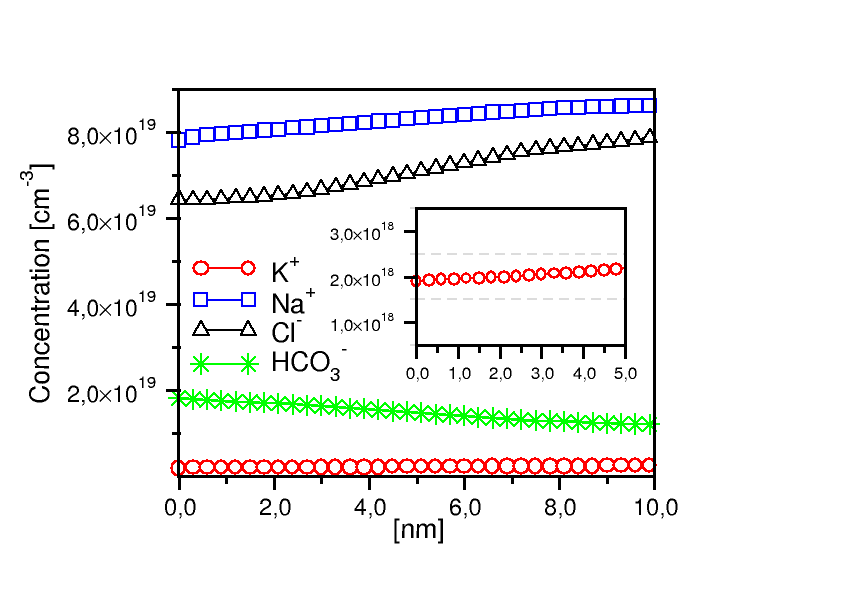}
		\caption{Sodium-potassium pump.}
		\label{fig:conc_KNa_pump}
	\end{subfigure}
	\begin{subfigure}[b]{0.495\textwidth}
		\centering
		\includegraphics[width = \textwidth]{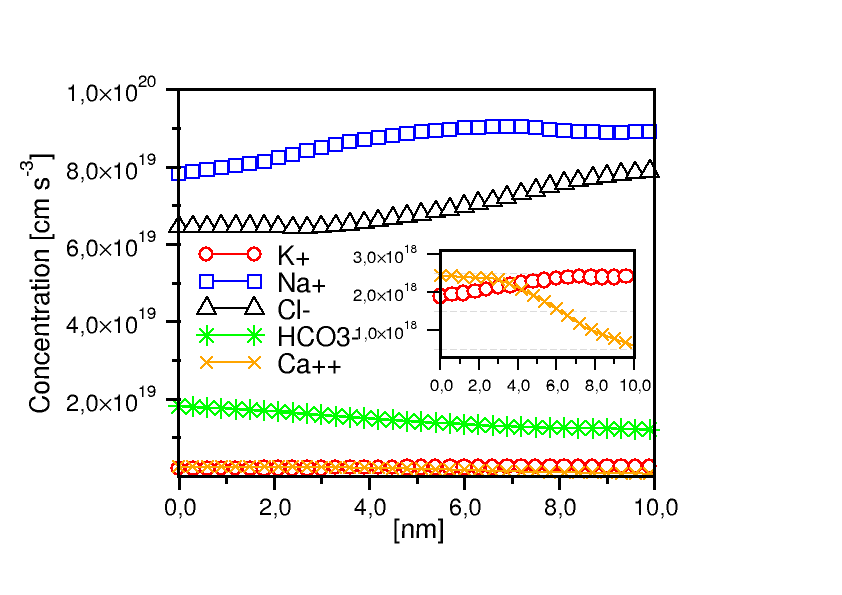}
		\caption{Calcium-sodium pump.}
		\label{fig:conc_CaNa_pump}
	\end{subfigure}

	\begin{subfigure}[b]{0.495\textwidth}
		\includegraphics[width = \textwidth]{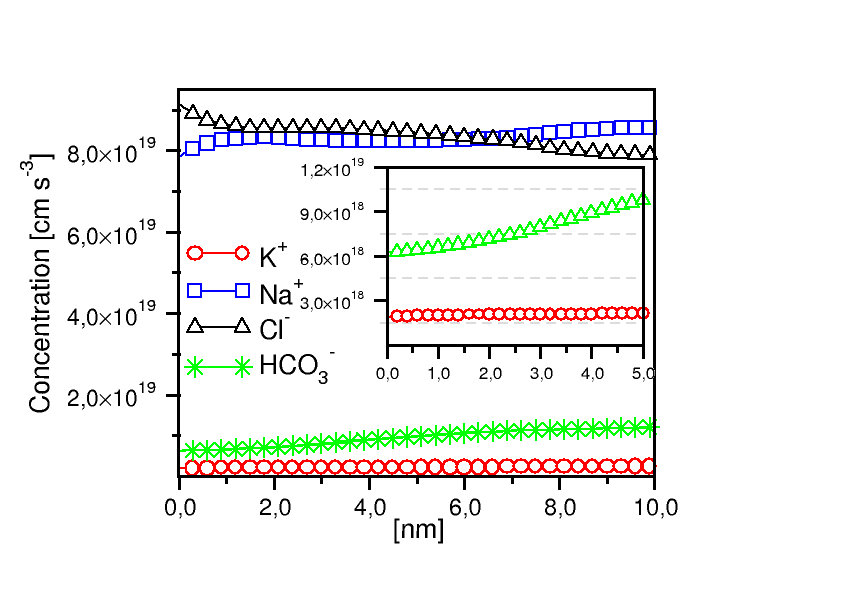}
		\caption{Anion channel.}
		\label{fig:conc_anion_channel}
	\end{subfigure}
	\begin{subfigure}[b]{0.495\textwidth}
		\includegraphics[width = \textwidth]{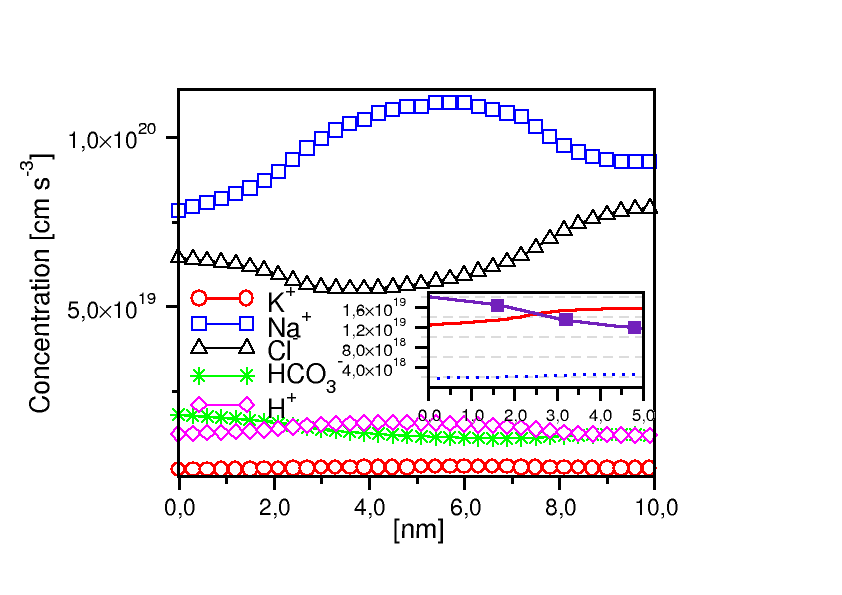}
		\caption{Hydrogenate-sodium pump.}
		\label{fig:conc_hydrogenate_sodium_pump}
	\end{subfigure}
	\caption{Computed ionic concentration along the $Z$ axis}
\end{figure}

\noindent
The computed profiles of the ion concentrations for each simulated ionic pump and channel are reported in 
Figures~\ref{fig:conc_KNa_pump}-~\ref{fig:conc_hydrogenate_sodium_pump}.
Results show the onset of a concentration gradient for each simulated ion species which appears not to be spatially constant 
for all ion species because of the action of the electric drift force which displaces the ion profile from the linear 
equilibrium distribution corresponding to a null electric field. 
Particularly worth noticing is the occurrence of significant variations for the concentration of the sodium ion in 
the simulation of the hydrogenate-sodium pump shown in Figure~\ref{fig:conc_hydrogenate_sodium_pump}.
These variations are the result of the attractive electrostatic force exerted on 
the sodium ions by the elevated negative 
fixed charge density distributed on the lateral surface on the channel region (cf.~Table~\ref{tab:sigma_fixed}).

\begin{figure}[h!]
		\centering
	\begin{subfigure}[b]{0.495\textwidth}
		\includegraphics[width = \textwidth]{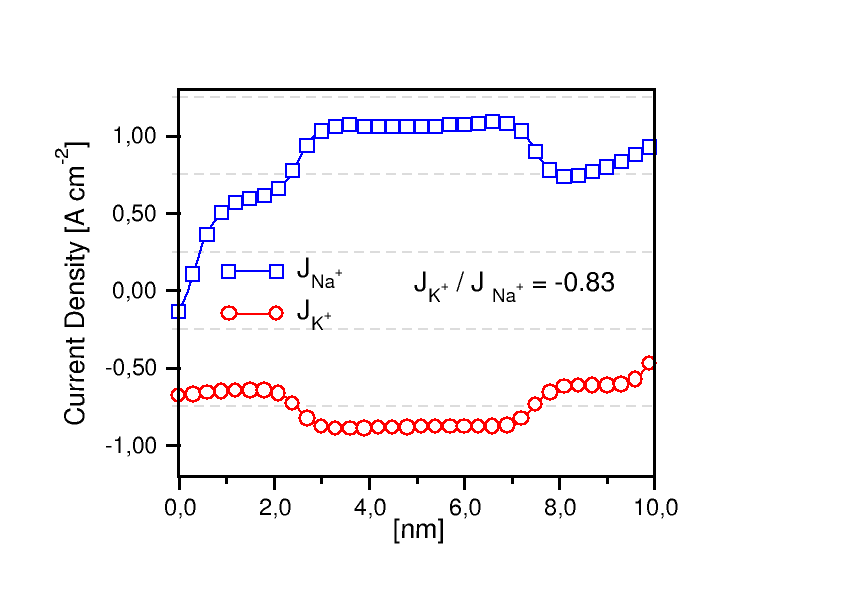}
		\caption{Sodium-potassium pump.}
		\label{fig:mech_CurrDens_K_Na}
	\end{subfigure}
	\begin{subfigure}[b]{0.495\textwidth}
		\includegraphics[width = \textwidth]{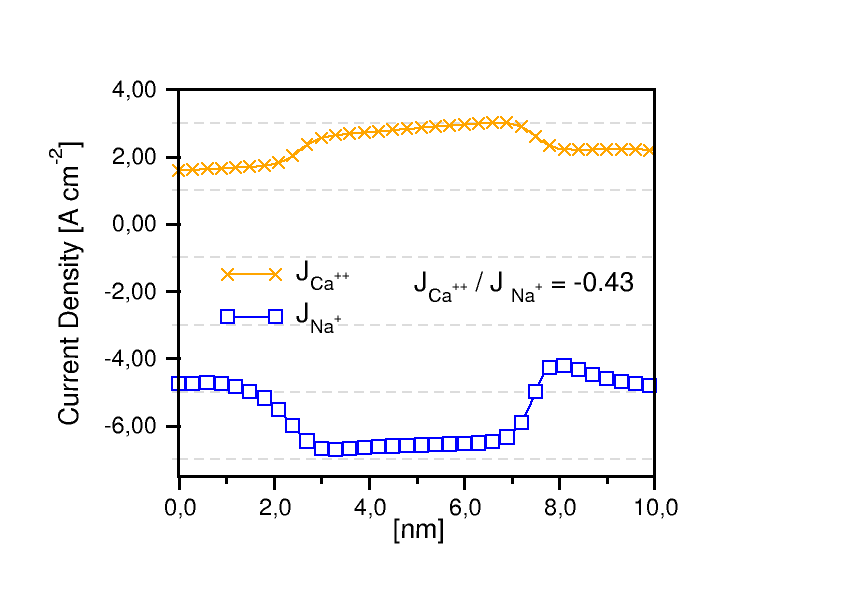}
		\caption{Calcium-sodium pump.}
		\label{fig:mech_CurrDens_Ca_Na}
	\end{subfigure}
	
	\begin{subfigure}[b]{0.495\textwidth}
		\includegraphics[width = \textwidth]{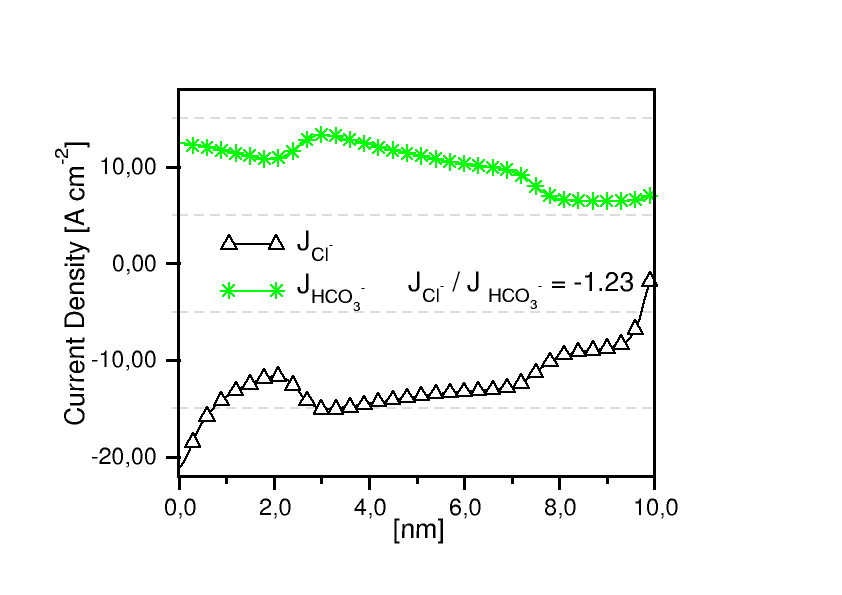}
		\caption{Anion channel.}
		\label{fig:mech_CurrDens_Cl_HCO3}
	\end{subfigure}
	\begin{subfigure}[b]{0.495\textwidth}
		\includegraphics[width = \textwidth]{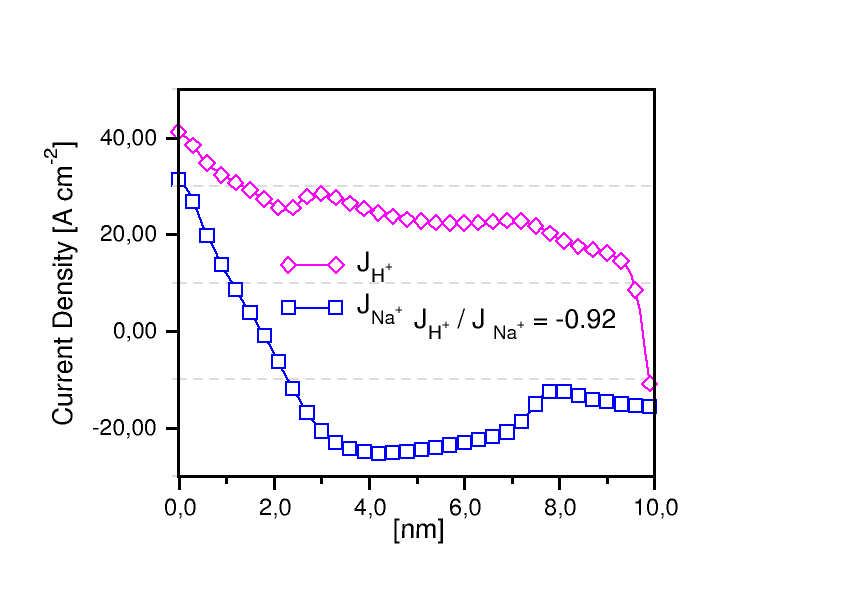}
		\caption{Hydrogenate-sodium pump.}
		\label{fig:mech_CurrDens_H_Na}
	\end{subfigure}
	\caption{Computed ionic current density along the $Z$ axis}
\end{figure}

Figures~\ref{fig:mech_CurrDens_K_Na}-~\ref{fig:mech_CurrDens_H_Na} show the spatial distributions of the computed
axial component of the ionic current density for each pump and channel. 
{For sake of clarity, in these} figures we report only the current
density related to the pump/exchanger functionality.
{We notice that for each ion, 
the value of the current density along the $Z$ axis is not constant because the cross-section
varies along the channel axis.} 

\noindent
Firstly, the sign of the computed current density for each simulated pump agrees with the theoretically expected direction. 
As a second comment, \textbf{\textit{the predicted values of the stoichiometric ratio $\mathcal{R}$ reasonably agrees with the corresponding 
theoretical value $r$}}.
Specifically, in the case of the sodium-potassium pump shown in Figure~\ref{fig:mech_CurrDens_K_Na}, 
$\mathcal{R}=0.83$, whereas $r=2:3=0.67$, in the case of the calcium-sodium pump shown in Figure~\ref{fig:mech_CurrDens_Ca_Na}, 
$\mathcal{R}=0.43$ whereas $r=1:3=0.33$, in the case of the anion channel shown in Figure~\ref{fig:mech_CurrDens_Cl_HCO3} 
we have $\mathcal{R}=1.23$ to be compared with $r=1:1$ and in the case of the hydrogenate-sodium pump shown 
in Figure~\ref{fig:mech_CurrDens_H_Na}) the value $\mathcal{R}=0.92$ fairly well agrees with the theoretical value $r=1:1$. 

\subsubsection{Fluid variables}
\begin{figure}[h!]
		\centering
		\includegraphics[width = 0.65\textwidth]{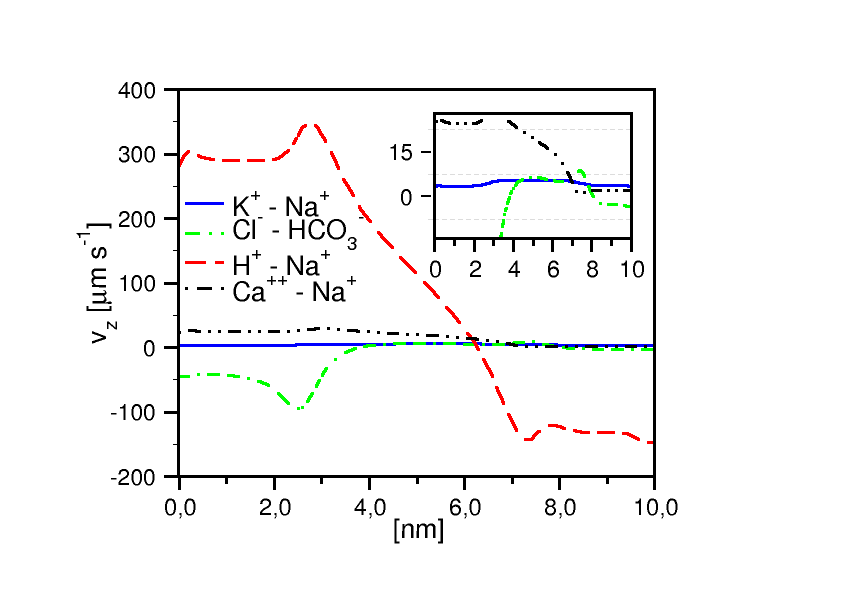}
		\caption{Computed spatial distribution of the axial component of fluid velocity for each pump/channel involved in AH production.}
		\label{fig:mech_vel1D}
\end{figure}

\noindent
Figure~\ref{fig:mech_vel1D} shows the spatial distribution of the axial component of the intrachannel water fluid velocity 
predicted by the simulation with the VE-PNP model of each ionic pump and channel involved in the process of AH production.
Results exhibit a significant difference among the various pumps mainly due to the presence of the fixed surface charge 
density $\sigma_{fixed}$ in the central region of the domain that {needed to be 
included to reproduce the correct pump functionalities.}
Specifically, in the case of the cation-based ionic pumps, model simulation predicts a positive value of the velocity 
in the whole computational domain whereas in the case of the anion channel the computed AH fluid velocity is strictly positive 
only in the central region of the domain.
In the case of the hydrogenate-sodium pump, the elevated fixed surface charge density on $\Gamma_{LAT_{INT}}$ gives rise to a 
change of sign of the axial component of the electric field at about the center of the domain. 
This, in turn, reflects into a change of sign in the volume force density $\uF_{ion}$ in the momentum balance 
equation of the fluid, causing an inversion of the intrachannel fluid flow at $Z=7.5 [\unit{nm}]$ where the direction 
of the axial velocity changes sign, from positive to negative.
{These results seem to suggest 
that the main ionic pump/exchangers contributing to AH secretion are those that actively 
involve Na$^{+}$.} 

\begin{figure}[h!]
	\centering
	\includegraphics[width = 0.85\textwidth]{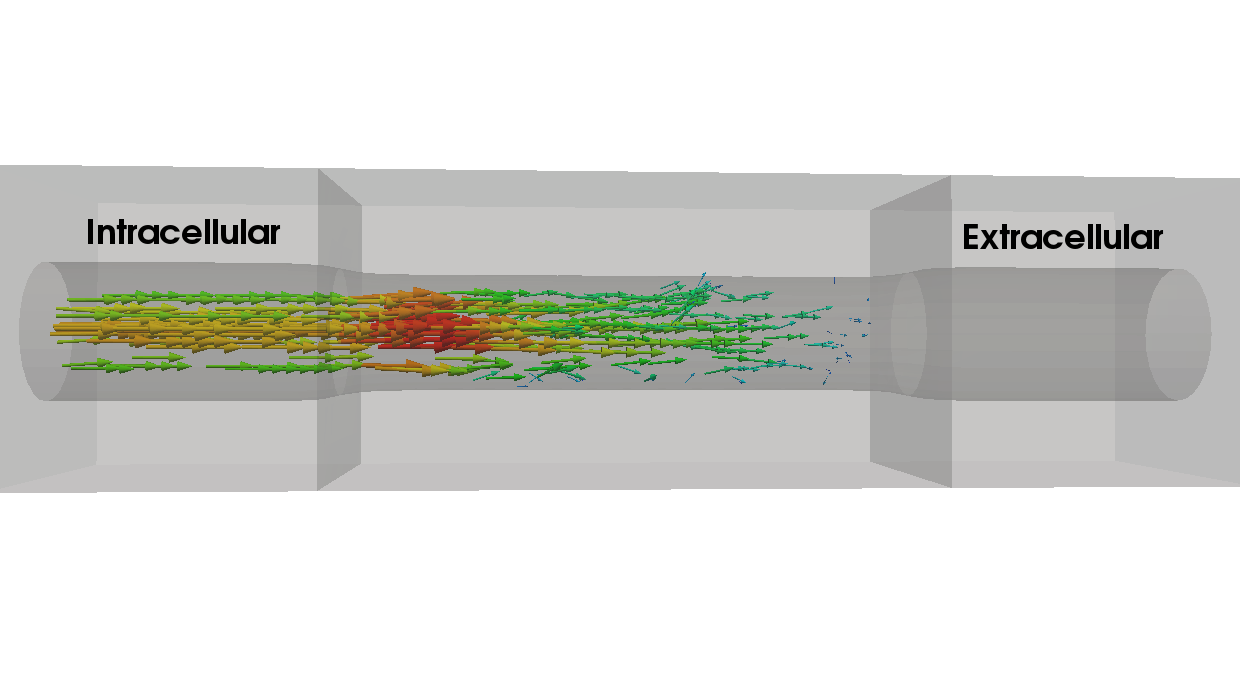}
	\caption{{Computed {three-dimensional} spatial distribution of fluid velocity in the case of the 
	{calcium-sodium} pump.}}
	\label{fig:mech_vel3D}
\end{figure}

Figure~\ref{fig:mech_vel3D} shows an example of {three-dimensional} computed spatial distribution of the
fluid velocity in the case of the {calcium-sodium} pump. 
{Results clearly show that AH flows from the intracellular space towards the
extracellular space.}

\subsubsection{Further remarks on pump functionality}
\begin{figure}[h!]
		\centering
	\begin{subfigure}[b]{0.495\textwidth}
		\includegraphics[width = \textwidth]{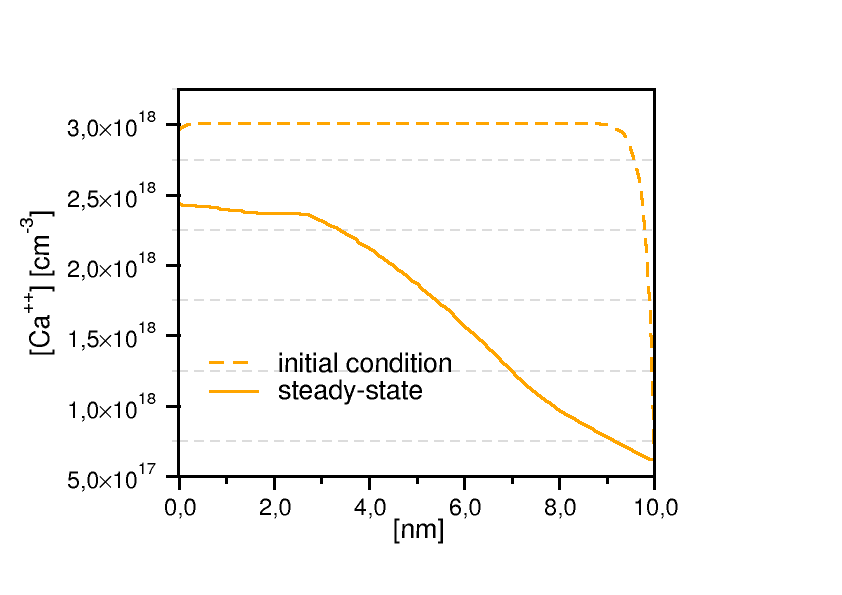}
		\caption{[Ca$^{++}$] in the calcium-sodium pump.}
		\label{fig:mech_confronto_Ca}
	\end{subfigure}
	\begin{subfigure}[b]{0.495\textwidth}
		\includegraphics[width = \textwidth]{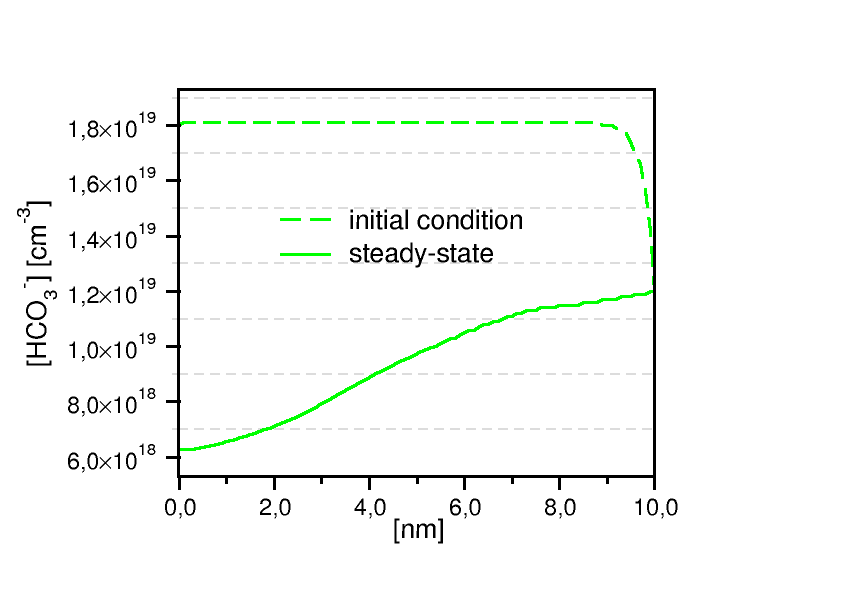}
		\caption{[HCO$_3^-$] for the anion channel.}
		\label{fig:mech_confronto_HCO3}
	\end{subfigure}
	
	\begin{subfigure}[b]{0.495\textwidth}
		\includegraphics[width = \textwidth]{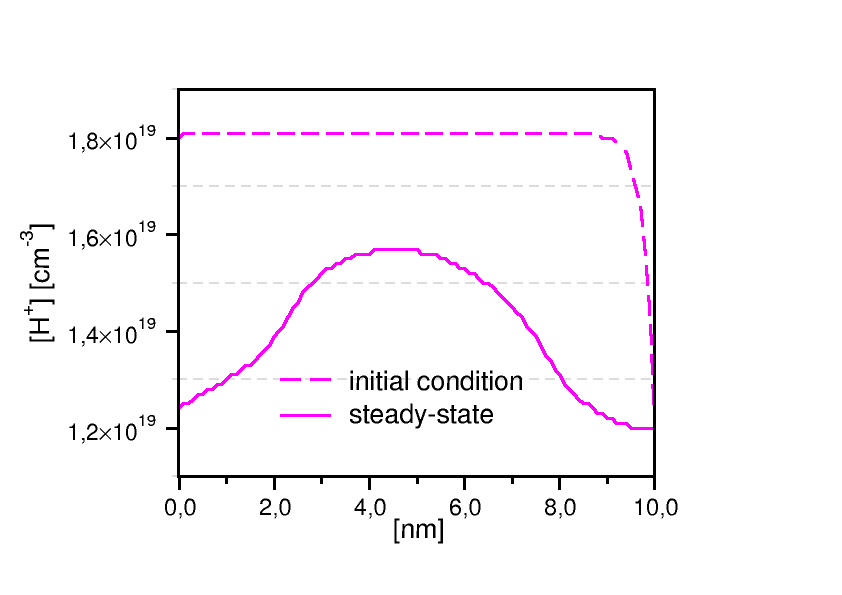}
		\caption{[H$^+$] for the hydrogenate-sodium pump}
		\label{fig:mech_confronto_H}
	\end{subfigure}
	\caption{Computed ionic current density along the $Z$ axis t various time:
		Dashed line: distribution at $t=0$. 
		Solid line: distribution at $t=T_{obs}$}
\end{figure}

In this section we briefly address a series of further considerations on the analysis of the simulation 
of the various ionic pumps and channels involved in the process of AH production, 
{in particular, those 
related to pump/exchanger functionality.}
The first consideration concerns the {temporal evolution of
calcium} in the Ca$^{++}$-Na$^+$ pump. 
To this purpose, Figure~\ref{fig:mech_confronto_Ca} illustrates the spatial calcium concentration at $t=0$ (dashed line)
and that at $t=T_{obs}$ (solid line). Results show a decrease of the level of calcium in the intracellular side of the domain. 
Such a decrease is slow because of the relatively low value $D_{\textrm{Ca}^{++}}=7.92 \cdot 10^{-6} \; [\unit{cm^2 s^{-1}}]$ 
of the diffusion coefficient adopted in the numerical simulation but, nonetheless, agrees with the theoretical expectation 
that physiological intracellular calcium level should be very low in healthy conditions (see~\cite{alberts2008molecular,ermentrout}).

\noindent

The second consideration concerns the time behavior of carbonate (HCO$_3^-$) in the anion channel. 
To this purpose, Figure~\ref{fig:mech_confronto_HCO3} illustrates the spatial carbonate concentration at $t=0$ (dashed line) 
and that at $t=T_{obs}$ (solid line). 
Results show a significant decrease of the level of carbonate in the intracellular region. 
This behavior agrees with the fact that the simulated electric field profile for the anion channel 
is positive in the intracellular region of the domain (cf.~Figure~\ref{fig:mech_Efield}) and therefore 
carbonate ions are swept away from left to right.
For further analysis of the importance of the carbonate ion in AH production, we refer to~\cite{Mauri2016}. 

The third consideration concerns the spatial distribution of hydrogenate (H$^+$) in the simulation of the hydrogenate-sodium pump. 
To this purpose, Figure~\ref{fig:mech_confronto_H} illustrates the spatial carbonate profile at $t=0$ (dashed line) and that 
at $t=T_{obs}$ (solid line).
Similarly to the previous case of carbonate, results show a significant reduction with simulation time of hydrogenate 
concentration in all the domain. 
However, unlike the previous case, concentration reduction equally occurs in both intracellular and extracellular sides whereas
hydrogenate accumulation occurs in the channel region. 
This behavior is due to the direction of the electric field (cf.~Figure~\ref{fig:mech_Efield}) which pushes H$^+$ from left to 
right in the intracellular side (where $E_Z >0$) and from right to left in the extracellular side (where $E_Z <0$).

\subsection{{Summary of the simulation results}}
In Table~\ref{tab:mech_results} we report the main outcomes of the simulation of ionic pumps and channels carried out 
in the context of AH secretion induced at the cellular scale level by the effect of ionic pressure
exerted on transmembrane fluid.
To summarize: \textbf{\textit{a unified modeling and computational framework allowed us to successfully simulate
the functionality of several ionic pump/exchangers}} {while preserving at the same time 
the features of each single exchanger, by a proper selection of the ion flux density BCs, of the osmotic 
gradient coefficient and of the amount of aminoacid charge in the channel protein folder}.
{These conclusions are significant outcomes of our computational model because 
osmotic gradient coefficient and permanent electric surface charge do not have yet a quantitative 
comparison with experimental data, though, they have been shown to be essential parts of
the biophysical description of the channel and to play a relevant role in determining AH flow direction.}

\begin{table}[h!]
\centering
\begin{tabular}{{|l|l|l|l|l|l|}}
\hline
		\small Pump & \small $k [\unit{N \, cm}]$ &\small $\sigma_{fixed} [\unit{C cm^{-2}}]$ 
		&\small  $\varphi_m [\unit{mV}]$ &\small  $\mathcal{R}$  &\small  $v_Z [\unit{\mu m \; s^{-1}}]$\\
		\hline \hline
		\scriptsize K$^+$ - Na$^+$ &\footnotesize $4.1\cdot10^{-19}$ &\footnotesize $-1.0\cdot10^{10}$ 
		&\footnotesize $-2.67$ &\footnotesize $0.83 (0.67)$ & $[4, \; 5.56]$ \\ \hline
		\scriptsize Cl$^-$ - HCO$_3^-$ &\footnotesize $ 24\cdot10^{-19} $ & 
		\footnotesize $+3.9\cdot 10^{11} $ &\footnotesize $-2.46 $ & \footnotesize$1.23 (1)$ & 
		$[-100, \; +6.56]$ \\ \hline
		\scriptsize H$^+$ - Na$^+$ &\footnotesize $ 4.0 \cdot10^{-19} $ & 
		\footnotesize $- 2.65 \cdot 10^{12} $ & \footnotesize$ -2.49 $ &\footnotesize $0.92 (1)$ & \footnotesize 
		$[-150, \; 300]$ \\ \hline
		\scriptsize Ca$^{++}$ - Na$^+$ &\footnotesize $3.95\cdot10^{-19} $ &
		\footnotesize $-6\cdot 10^{11} $ &\footnotesize $-2.39 $ & \footnotesize$0.43 (0.33)$ &\footnotesize 
		$[2, \; 28]$ \\	\hline
	\end{tabular}
	\vspace*{4pt}
	\caption{A summary of the simulation results for ionic pumps and channel involved in AH production. 
		Aside the predicted value of $\mathcal{R}$ we report in braces the theoretically expected value.
		The column $v_Z$ reports for each row the predicted range of the fluid velocity.}
	\label{tab:mech_results}
\end{table}

\section{Conclusions, model limitations and future objectives}\label{sec:final_conclusions}
A unified modeling and computational framework with {electrochemical osmotic correction}
and with a realistic geometry to represent the computational domain, 
has been proposed to investigate the main functional principles of the sodium-potassium pump, 
the calcium-sodium pump, the anion channel and the hydrogenate-sodium pump, that are involved in the
production of aqueoud humor
in the ciliary body of the eye.
{The theoretical model has been demonstrated to correctly reproduce, for each simulated ionic pump and  
channel, existing experimental data of transepithelial membrane potential in animal models.
The model has also allowed, for the first time to the best of our knowledge in the study 
of AH production, the quantitative analysis of novel biophysical mechanisms 
such as the physiological stoichiometric rate, the direction of AH flow and the magnitude of AH fluid velocity.
Thus, the present study motivates the further development of this modeling approach to 
\textit{(i)} simulate the simultaneous presence and action of the several ionic pump/exchangers 
 considered in this work, with the aim of quantitatively estimating their reciprocal influence;
\textit{(ii)} include the presence of other molecules actively transported through the cell membrane, including the ascorbic acid, 
 which is secreted by a transporter (Sodium-dependent vitamin C transporter 2 (SVCT2)) \cite{Tsukaguchi1999}; and 
\textit{(iii)} the simulation of {the effect of administration of a drug} in the regulation of AH secretion.

It is expected that such theoretical advancement of the frontier of knowledge in this branch of Human Sciences
may significantly help design new molecules for drug synthesis and, as a consequence, considerably 
reduce time and costs for clinical availability of new pharmacological therapies.

It is important to emphasize, though, that a number of biophysical limitations still affect the proposed mathematical model 
of aqueous humor dynamics. Among them, we mention that our model does not account for \textit{(i)} autonomic system pathways, specifically, 
the sympathetic and parasympathetic (see~\cite{Enciclopediaoftheeye_2010}); 
\textit{(ii)} variations due to the circadian rhythm (see~\cite{circadianAH});
\textit{(iii)} the role of aquaporins in the exchange of fluid across the cell membrane (see~\cite{aquaporinsAH}). A research effort to address these limitations is currently in progress.}

\section*{Acknowledgements}
PhD candidate Sala is supported by scholarship of Ministère de l’Enseignement supérieur et de la 
Recherche (France). Dr. Sacco has been partially supported by Micron Semiconductor Italia S.r.l., 
SOW nr. 4505462139. Dr. Guidoboni has been partially supported by the award 
NSF DMS-1224195, the Chair Gutenberg funds of the
Cercle Gutenberg (France) and the LabEx IRMIA (University of Strasbourg, France). 
Dr. Harris has been partially supported by Research to Prevent Blindness (RPB, NY, USA)
\section*{Data tables}
\begin{table}[h!]
	\centering
	\begin{tabular}{{|l|l|l|}}
		\hline
		K$^+_0$ & $g_{\textrm{K}^+}$ & K$^+_{out}$ \\	\hline
		$2.41 \cdot 10^{18} \; [\unit{cm^{-3}}]$ & $+4 \cdot 10^{17} \; [\unit{cm^{-2} s^{-1}}]$& 
		$2.41 \cdot 10^{18} \; [\unit{cm^{-3}}]$ \\ \hline \hline
		Na$^+_0$ & Na$^+_{in}$ & $g_{\textrm{Na}^+}$ \\ \hline
		$8.19 \cdot 10^{19} \; [\unit{cm^{-3}}]$ & 
		$7.82 \cdot 10^{19} \; [\unit{cm^{-3}}]$ & $+6 \cdot 10^{17} \; [\unit{cm^{-2} s^{-1}}]$ \\ \hline
	\end{tabular}
	\caption{Boundary and initial data for the cations in Sec. \ref{sec:simulations_1}.}
	\label{tab:Ion+}
\end{table}

\begin{table}[h!]
	\centering
	\begin{tabular}{{|l|l|l|}}
		\hline
		Cl$^-_0$ & Cl$^-_{in}$ & Cl$^-_{out}$\\	\hline
		$7.17 \cdot 10^{19} \; [\unit{cm^{-3}}]$ & 
		$6.44 \cdot 10^{19} \; [\unit{cm^{-3}}]$ & 
		$7.89 \cdot 10^{19} \; [\unit{cm^{-3}}]$ \\ \hline \hline
		HCO$^{-}_{3,0}$ & HCO$^{-}_{3,in}$ & HCO$^{-}_{3,out}$ \\ \hline
		$1.51 \cdot 10^{19} \; [\unit{cm^{-3}}]$ & $1.81 \cdot 10^{19} \; [\unit{cm^{-3}}]$ & 
		$1.2 \cdot 10^{19} \; [\unit{cm^{-3}}]$ \\ \hline
	\end{tabular}
	\caption{Boundary and initial data for the anions in Sec. \ref{sec:simulations_1}.}
	\label{tab:Ion-}
\end{table}


\begin{table}[h!]
	\begin{tabular}{{|l|l|l|}}
		\hline
		K$^+_0$ $[\unit{cm^{-3}}]$ & $g_{\textrm{K}^+}$ $[\unit{cm^{-2} s^{-1}}]$ & K$^+_{out}$ $[\unit{cm^{-3}}]$\\ \hline
		$2.41\cdot10^{18}$ & $-4\cdot10^{19}$ & $2.41\cdot10^{18}$ \\ \hline \hline
		Na$^+_0$ $[\unit{cm^{-3}}]$ & Na$^+_{in}$ $[\unit{cm^{-3}}]$ & $g_{\textrm{Na}^+}$ 
		$[\unit{cm^{-2} s^{-1}}]$ \\ \hline
		$8.19\cdot10^{19}$ & $7.82\cdot10^{19}$ & $-6\cdot10^{19}$ \\ \hline \hline
		Cl$^-_0$ $[\unit{cm^{-3}}]$ & Cl$^-_{in}$ $[\unit{cm^{-3}}]$ & Cl$^-_{out}$ $[\unit{cm^{-3}}]$\\ \hline			
		$7.17\cdot10^{19}$ & $6.44\cdot10^{19}$ & $7.89\cdot10^{19}$ \\ \hline	\hline
		HCO$_3^-$$_0$ $[\unit{cm^{-3}}]$ & HCO$_3^-$$_{in}$ $[\unit{cm^{-3}}]$ & 
		HCO$_3^-$$_{out}$ $[\unit{cm^{-3}}]$ \\ \hline			
		$1.51\cdot10^{19}$ & $1.81\cdot10^{19}$ & $1.20\cdot10^{19}$ \\ \hline
	\end{tabular}
	\caption{Data for the sodium-potassium pump in Sec. \ref{sec:simulations_2}.}
	\label{tab:sp_pump_2}
\end{table}

\begin{table}[h!]
	\begin{tabular}{{|l|l|l|}}
		\hline
		K$^+_0$ $[\unit{cm^{-3}}]$ & K$^+_{in}$ $[\unit{cm^{-3}}]$ & K$^+_{out}$ $[\unit{cm^{-3}}]$ \\ \hline
		$2.41\cdot10^{18}$ & $1.90\cdot10^{18}$ & $2.41\cdot10^{18}$ \\ \hline \hline
		Na$^+_0$ $[\unit{cm^{-3}}]$ & Na$^+_{in}$ $[\unit{cm^{-3}}]$ & $g_{\textrm{Na}^+}$ 
		$[\unit{cm^{-2} s^{-1}}]$ 
		\\ \hline
		$8.19\cdot10^{19}$ & $7.82\cdot10^{19}$ & $6\cdot10^{19}$ \\ \hline	\hline
		Ca$^{++}_0$ $[\unit{cm^{-3}}]$ & $g_{\textrm{Ca}^{++}}$ $[\unit{cm^{-2} s^{-1}}]$ & Ca$^{++}_{out}$ 
		$[\unit{cm^{-3}}]$ \\ \hline
		$3.011\cdot10^{18}$ & $2\cdot10^{19}$ & $6.022\cdot10^{17}$ \\ \hline	\hline
		Cl$^-_0$ $[\unit{cm^{-3}}]$ & Cl$^-_{in}$ $[\unit{cm^{-3}}]$ & Cl$^-_{out}$ $[\unit{cm^{-3}}]$ 
		\\ \hline			
		$7.17\cdot10^{19}$ & $6.44\cdot10^{19}$ & $7.89\cdot10^{19}$ \\ \hline	\hline
		HCO$_3^{-}$$_0$ $[\unit{cm^{-3}}]$ & HCO$_3^{-}$$_{in}$ $[\unit{cm^{-3}}]$ 
		& HCO$_3^{-}$$_{out}$ $[\unit{cm^{-3}}]$\\ \hline			
		$1.51\cdot10^{19}$ & $1.81\cdot10^{19}$ & $1.2\cdot10^{19}$ \\ 	\hline
	\end{tabular}
	\caption{Data for the calcium-sodium pump in Sec. \ref{sec:simulations_2}.}		
	\label{tab:cas_pump_2}
\end{table}

\begin{table}[h!]
	\begin{tabular}{{|l|l|l|}}
		\hline
		K$^+_0$ $[\unit{cm^{-3}}]$ & K$^+$$_{in}$ $[\unit{cm^{-3}}]$ & K$^+$$_{out}$ $[\unit{cm^{-3}}]$ 
		\\ \hline
		$2.41\cdot10^{18}$ & $1.90\cdot10^{18}$ & $2.41\cdot10^{18}$ \\ \hline\hline
		Na$^+_0$ $[\unit{cm^{-3}}]$ & Na$^+_{in}$ $[\unit{cm^{-3}}]$ & Na$^+$$_{out}$ $[\unit{cm^{-3}}]$ 
		\\ \hline
		$8.19\cdot10^{19}$ & $7.82\cdot10^{19}$ & $8.55\cdot10^{19}$ \\	\hline	\hline
		Cl$^-$$_0$ $[\unit{cm^{-3}}]$ & $g_{\textrm{Cl}^-}$ $[\unit{cm^{-2} s^{-1}}]$ & 
		Cl$^-$$_{out}$ $[\unit{cm^{-3}}]$\\ \hline 			
		$7.17\cdot10^{19}$ & $8\cdot10^{19}$ & $7.89\cdot10^{19}$ \\ \hline	\hline
		HCO$_3^{-}$$_0$ $[\unit{cm^{-3}}]$ & $g_{\textrm{HCO}_3^-}$ $[\unit{cm^{-2} s^{-1}}]$ & 
		HCO$_3^{-}$$_{out}$ $[\unit{cm^{-3}}]$\\ \hline 			
		$1.81\cdot10^{19}$ & $-8\cdot10^{19}$ & $1.20\cdot10^{19}$ \\	\hline
	\end{tabular}
	\caption{Data for the anion channel in Sec. \ref{sec:simulations_2}.}
	\label{tab:anion_ch_2}	
\end{table}

\begin{table}[h!]
	\begin{tabular}{{|l|l|l|}}
		\hline
		K$^+_0$ $[\unit{cm^{-3}}]$ & K$^+_{in}$ $[\unit{cm^{-3}}]$ & 
		$K^+_{out}$ $[\unit{cm^{-3}}]$\\ \hline
		$2.41\cdot10^{18}$ & $1.90\cdot10^{18}$ & $2.41\cdot10^{18}$ \\ \hline \hline
		Na$^+_0$ $[\unit{cm^{-3}}]$ & Na$^+_{in}$ $[\unit{cm^{-3}}]$ & $g_{\textrm{Na}^+}$ 
		$[\unit{cm^{-2} s^{-1}}]$\\ \hline
		$8.19\cdot10^{19}$ & $7.82\cdot10^{19}$ & $1.0\cdot10^{20}$ \\ 	\hline \hline		
		H$^{+}_0$ $[\unit{cm^{-3}}]$ & $g_{\textrm{H}^{+}}$ $[\unit{cm^{-2} s^{-1}}]$ & H$^{+}_{out}$ 
		$[\unit{cm^{-3}}]$\\ \hline
		$1.81\cdot10^{19}$ & $1.0\cdot10^{20}$ & $1.20\cdot10^{19}$ \\ \hline \hline
		Cl$^-_0$ $[\unit{cm^{-3}}]$ & Cl$^-_{in}$ $[\unit{cm^{-3}}]$ & Cl$^-_{out}$ $[\unit{cm^{-3}}]$
		\\ \hline 			
		$7.17\cdot10^{19}$ & $6.44\cdot10^{19}$ & $7.89\cdot10^{19}$ \\	\hline \hline
		HCO$_3^{-}$$_0$ $[\unit{cm^{-3}}]$ & HCO$_3^{-}$$_{in}$ $[\unit{cm^{-3}}]$ 
		& HCO$_3^{-}$$_{out}$ $[\unit{cm^{-3}}]$\\ \hline 			
		$1.81\cdot10^{19}$ & $1.81\cdot10^{19}$ & $1.20\cdot10^{19}$ \\ \hline
	\end{tabular}
	\caption{Data for the hydrogenate-sodium pump in Sec. \ref{sec:simulations_2}.}
	\label{tab:hyd_s_pump_2}
\end{table}

%

\end{document}